\documentclass[letter, superscriptaddress, prd, twocolumn, showpacs, nofootinbib]{revtex4}

\usepackage{verbatim, amsmath, amssymb, graphicx, color}

\begin{document}

\title{Evolution of curvature and anisotropy near a nonsingular bounce}

\date{}

\author{BingKan Xue}
\affiliation{Department of Physics, Princeton University, Princeton,
New Jersey 08544, USA}

\author{Paul J. Steinhardt}
\affiliation{Department of Physics, Princeton University, Princeton,
New Jersey 08544, USA}
\affiliation{Princeton Center for Theoretical
Physics, Princeton University, Princeton, New Jersey 08544, USA}

\begin{abstract}
We consider bouncing cosmologies in which an ekpyrotic contraction phase with $w \gg 1$ is followed by a bouncing phase with $w < -1$ that violates the null energy condition. The bouncing phase, induced by ghost condensation, is designed to produce a classically nonsingular bounce at a finite value of the scale factor. We show that the initial curvature and anisotropy, though diluted during the ekpyrotic phase, grow back exponentially during the bouncing phase. Moreover, curvature perturbations and anisotropy are generated by quantum fluctuations during the ekpyrotic phase. In the bouncing phase, however, an adiabatic curvature perturbation grows to dominate and contributes a blue spectrum that spoils the scale-invariance. Meanwhile, a scalar shear perturbation grows nonlinear and creates an overwhelming anisotropy that disrupts the nonsingular bounce altogether.
\end{abstract}


\maketitle

\section{Introduction}

Cosmological models are expected to explain the large scale properties of the early Universe: the homogeneity, flatness and isotropy of the background, and the nearly scale-invariant spectrum of the primordial perturbations. These conditions can be achieved in either a rapidly expanding (inflationary) phase right after the big bang \cite{Guth:1980zm, Linde:1981mu, Albrecht:1982wi} or a slow contracting (ekpyrotic) phase preceding the big bang \cite{Khoury:2001wf, Erickson:2003zm}. In order to have the latter scenario, however, the Universe has to undergo a bounce from the contracting phase to an expanding phase. This transition imposes an extra condition on cosmological models, namely, the desired properties of the large scale structure created in the contracting phase must be propagated safely through the bounce. The bounce may be induced by quantum gravity effects when the scale factor shrinks to the Planck scale near a classical singularity \cite{Turok:2004gb, Turok:2007ry, McFadden:2005mq}. Although theories of such a quantum bounce (also referred to as a ``singular'' bounce) are not fully developed, it is conjectured that the homogeneity and the scale-invariance conditions generated before the bounce should pass smoothly through the bounce \cite{Turok:2004gb, Turok:2007ry, McFadden:2005mq, Tolley:2003nx}. The intuition is that the low energy physics that generates structure on large length scales should decouple from the physics at high energies and short lengths that is responsible for the bounce.

An alternative approach that has been pursued in recent years is to realize the bounce in a classically nonsingular way \cite{Creminelli:2007aq, Buchbinder:2007ad, Lin:2010pf}. In this approach the Universe stops contraction and reverses to expansion at a finite value of the scale factor $a$, when classical gravity and effective field theories are still valid. The intended advantage is that it becomes possible to describe with known theories the entire evolution of the cosmological background and the conditions that emerge from the bounce. To have a nonsingular bounce requires violation of the null energy condition (NEC). A commonly used approach is to introduce a scalar field that undergoes {\it ghost condensation} \cite{ArkaniHamed:2003uy}. The period during which the NEC is violated will be referred to as the \emph{bouncing phase}.

Some earlier studies \cite{Lin:2010pf} of nonsingular bounces have not included an ekpyrotic phase. In those cases, the contracting universe is unstable to the growth of anisotropies unless one assumes extreme fine-tuning of initial conditions; it is also difficult to produce a sufficiently wide spectrum of scale-invariant fluctuations. Moreover, some earlier treatments \cite{Lin:2010pf, Creminelli:2007aq} of the bouncing phase only consider the last stage when the Hubble parameter $H$ varies linearly with time and passes through zero. This period does not cover the whole bouncing phase during which NEC is violated, leaving out important problems that arise in earlier stages of the bouncing phase.

In this paper, we consider nonsingular bouncing cosmologies that have an ekpyrotic phase prior to the bouncing phase. The ekpyrotic phase stably smooths and flattens the Universe, and leads to the generation of nearly scale-invariant perturbations. After the ekpyrotic phase, ghost condensation is triggered to bring the Universe into a bouncing phase with $w < -1$ that violates the NEC. We consider the entire bouncing phase from the start of NEC violation at $\dot{H} = 0$ to the nonsingular bounce at $H = 0$. We examine perturbations around the background evolution and identify problems that break the homogeneity, isotropy and scale-invariance conditions.

In a previous paper \cite{Xue:2010ux}, we argued that the bouncing phase strongly modifies the spectrum of curvature perturbations. An adiabatic mode, though decaying and negligible during the ekpyrotic phase, was shown to grow exponentially just before the bouncing phase when the equation of state $w$ crosses $-1$. By the time the Universe bounces, the total curvature perturbation becomes dominated by this adiabatic mode, which unfortunately carries a blue spectrum that is inconsistent with cosmological observations. This result does not depend on how the scale-invariant mode is generated in the ekpyrotic phase. To further elaborate this idea, we will show that our conclusion does not change with different gauge choices, hence posing a serious physical problem for nonsingular bouncing models.

In addition, we will demonstrate a new effect that significantly amplifies the anisotropy. This amplification occurs gradually through the entire bouncing phase, and is not noticed in previous analyses that focused only on the last stage of the bounce when $H$ increases linearly with time. In our analysis, we find that the scale factor has to decrease exponentially during the entire bouncing phase, causing the anisotropy to grow substantially before the bounce. Consequently, the homogeneity and isotropy of the Universe achieved in the preceding ekpyrotic phase are lost. Here we consider both anisotropies that have classical and quantum origins. The classical anisotropy, present at the beginning of the ekpyrotic phase, generally comes to dominate over the ghost condensate field before the bounce, unless it is tuned to a sufficiently small value even before the ekpyrotic phase begins. A more serious issue is the anisotropy induced by quantum fluctuations, whose amplitude is linked to the amplitude of curvature perturbations. This quantum generated anisotropy is negligible during the ekpyrotic phase, but becomes overwhelmingly large in the bouncing phase. As a result, the Universe is inevitably led to chaotic mixmaster behavior and contracts to a singularity, $a \rightarrow 0$; that is, the nonsingular bounce is totally disrupted.

In Sec.~\ref{sec:background}, we present a sample model that incorporates the generic features of an ekpyrotic phase and a nonsingular bouncing phase. The bouncing phase is described in detail in Sec.~\ref{sec:bounce}. Then in Sec.~\ref{sec:curvature}, we demonstrate the dramatic growth of curvature perturbation that alters the power spectrum, presenting our analysis in both the comoving and the synchronous gauges. In Sec.~\ref{sec:anisotropy}, we analyze the evolution of the anisotropy, showing how it is induced by the curvature perturbation and how it disrupts the nonsingular bounce. In Sec.~\ref{sec:conclusion} we argue that both problems of curvature and anisotropy can be attributed to the exponential difference between the low energy scale of ghost condensation and the high energy scale of the ekpyrotic phase.

\section{The ekpyrotic nonsingular bouncing model} \label{sec:background}

In this Section, we introduce a generic nonsingular bouncing cosmology that we will use for our study. Previous analyses of nonsingular bouncing models can be divided into those with an ekpyrotic contraction phase with equation of state $w > 1$ and those in which the contraction phase has $w$ strictly $<1$. Without an ekpyrotic phase, the contraction is unstable to the growth of anisotropy and the onset of chaotic mixmaster oscillations \cite{Erickson:2003zm}. To counter the instability, extreme fine-tuning of initial conditions must be imposed, or the unstable contracting phase must be made so short that it is insufficient to explain the large scale uniformity and density perturbations observed today. For these reasons, the only cases of interest are those with a long ekpyrotic contraction phase with $w > 1$, followed by a transition to a bouncing phase with $w < -1$.

The transition is hard to achieve by combining two independent components with $w > 1$ and $w < -1$ respectively, because if the $w > 1$ component dominates during the ekpyrotic phase, then the $w < -1$ component would grow much more slowly and never come to dominate during the bouncing phase. A more economical approach is to have a single component whose equation of state $w$ switches from $> 1$ to $< -1$ between the ekpyrotic and the bouncing phase. This scenario is achieved naturally by having a scalar field that rolls down a steep negative potential during the ekpyrotic phase, and then undergoes ghost condensation after the ekpyrotic phase is over. In this way the entire contracting and bouncing history of the Universe is driven by the same field.

\begin{figure}
\centering
\includegraphics[width=0.4\textwidth]{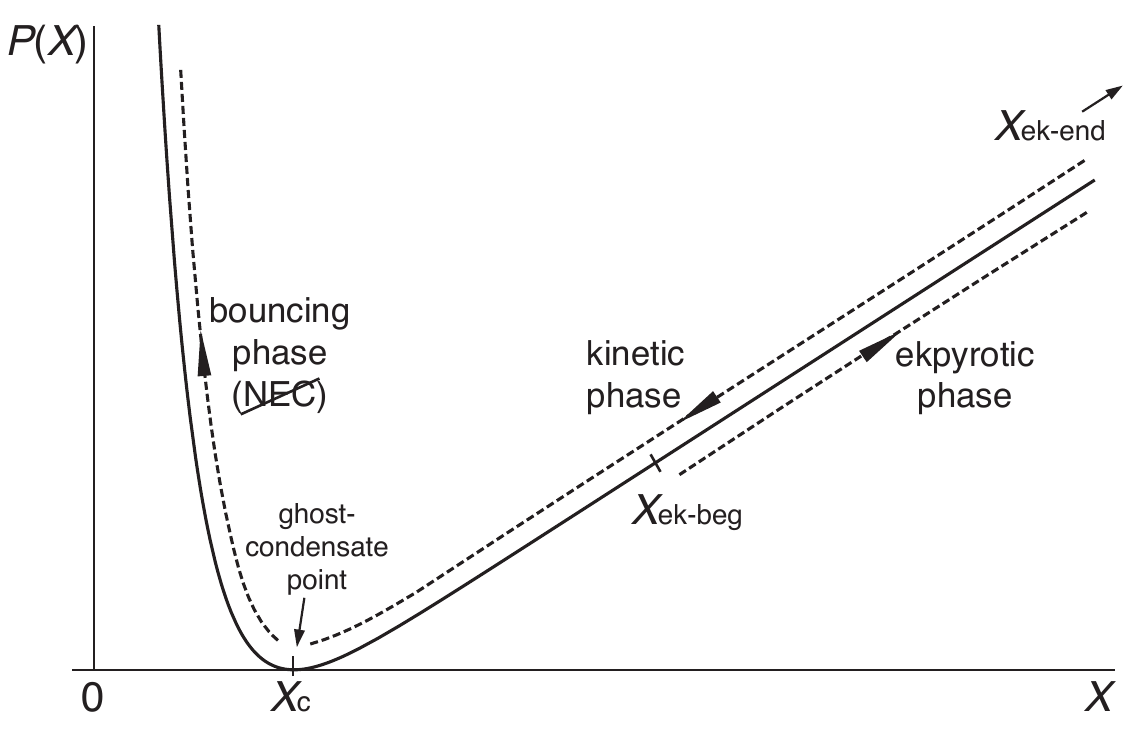}
\caption{The kinetic term $P(X)$ versus $X \equiv \frac{1}{2} (\partial \phi)^2$ in the effective Lagrangian for the scalar field $\phi$. During the ekpyrotic phase, $X$ is in the linear region $X \gg X_\text{c}$ and increases by a factor $e^{2N}$. In the transition to the bouncing phase, $X$ decreases by an even greater factor to reach $X = X_\text{c}$. During the bouncing phase, $X$ further decreases to $X < X_\text{c}$ where NEC is violated.}
\label{fig:kinetic}
\end{figure}

\begin{figure}
\centering
\includegraphics[width=0.4\textwidth]{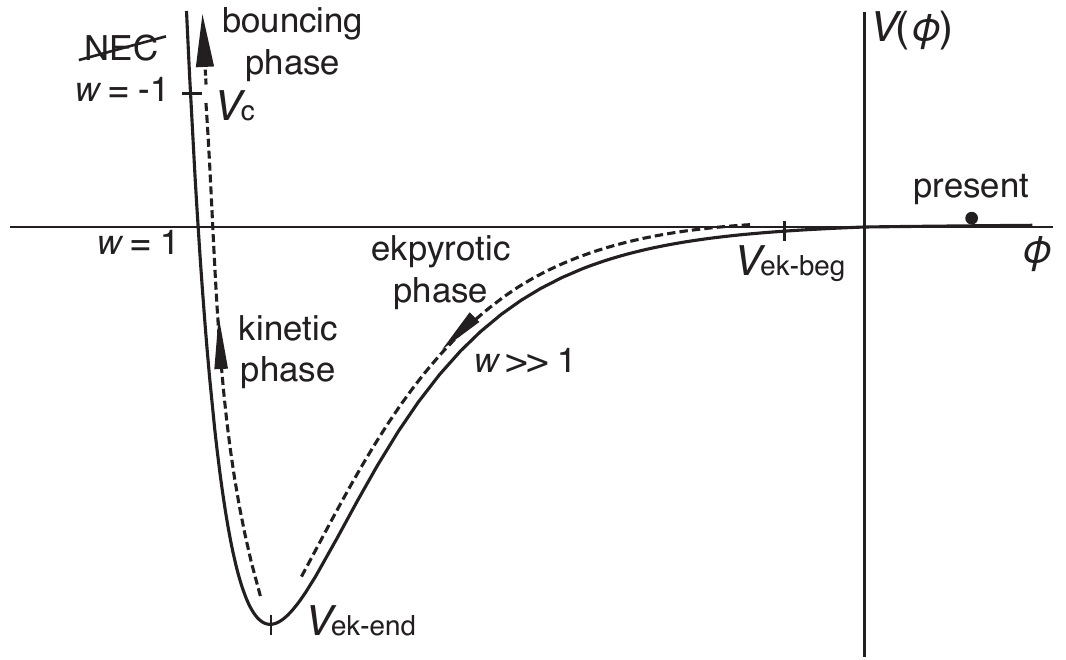}
\caption{The potential term $V(\phi)$ in the effective Lagrangian for the scalar field $\phi$. The ekpyrotic phase corresponds to the exponential decline from $V_\text{ek-beg}$ to $V_\text{ek-end}$ near the minimum of the potential. The kinetic phase refers to the quick rise from $V_\text{ek-end}$ to $V_\text{c} \approx 3p \, |V_\text{ek-end}|$. The nonsingular bouncing phase occurs at $V > V_\text{c}$.}
\label{fig:potential}
\end{figure}

This framework is described, for instance, in the new ekpyrotic model \cite{Buchbinder:2007ad}, where the effective Lagrangian for the scalar field $\phi$ is
\begin{equation} \label{eq:L}
\mathcal{L} = \sqrt{-g} \big[ P(X) - V(\phi) \big], \quad X \equiv - \tfrac{1}{2} \left( \partial \phi \right) ^2 .
\end{equation}
The background metric $g_{\mu\nu}$ is taken to be flat Friedmann-Robertson-Walker,
\begin{equation} \label{eq:FRW}
ds^2 = - dt^2 + a(t)^2 \big( dx^2 + dy^2 + dz^2 \big) ,
\end{equation}
and we use reduced Planck units $8 \pi G \equiv 1$. The kinetic term $P(X)$, shown in Fig.~\ref{fig:kinetic},
is canonical, $P(X) \approx X$, for large $X$; but has a minimum at a low energy scale $X_\text{c}$ where ghost condensation takes place. The potential $V(\phi)$ is shown in Fig.~\ref{fig:potential},
where, from right to left, it is first approximated by a negative exponential, $-V_0 \, e^{- \sqrt{2/p} \, \phi}$ with $p \ll 1$; then bottoms out and undergoes a steep rise. Unlike in \cite{Buchbinder:2007ad, Creminelli:2007aq} where $V$ is designed to bend sharply at a fine-tuned value above zero, we choose to let it rise smoothly as in Fig.~\ref{fig:potential}. The Universe evolves through the ekpyrotic phase ($w \gg 1$) to the bouncing phase ($w < -1$), with a transient kinetic energy dominated phase ($w \approx 1$) in between, as indicated in the figures.

The \emph{ekpyrotic phase} starts at an intermediate energy scale $X_\text{ek-beg}$ when the scalar field rolls down the potential from $V_\text{ek-beg}$, and ends when the field reaches the bottom of the potential $V_\text{ek-end}$. The negative exponential form of the potential leads to a homogeneous attractor solution,
\begin{align} \label{eq:attractor}
& a = \alpha (-t)^p, \quad \alpha \equiv a_\text{ek-end} (-H_\text{ek-end}/p)^p , \\
& H \equiv \frac{\dot{a}}{a} = - \frac{p}{(-t)} \; , \\
& \phi = \sqrt{2p} \, \log \Big( \sqrt{\tfrac{V_0}{p(1-3p)}} \, (-t) \Big),
\end{align}
where $t$ is negative and increasing towards zero. This solution has a constant equation of state,
\begin{equation}
w = \tfrac{2}{3p} - 1 \gg 1 \, ,
\end{equation}
where typically $p \sim 10^{-2}$ \cite{Khoury:2003rt}. There is a scaling relation between the kinetic and the potential energies,
\begin{equation} \label{eq:scaling}
H^2 = \frac{p}{3p-1} \, V = p \, X \, , \quad X = \frac{1}{2} \dot{\phi}^2 .
\end{equation}
According to the first Friedmann equation, we have
\begin{equation} \label{eq:Friedmann1}
H^2 = \frac{1}{3} \Big( -\frac{3{\rm k}}{a^2} + \frac{\sigma_0^2}{a^6} + \frac{{\rho_\phi}_0}{a^{3 (1 + w)}} + \cdots \Big) ,
\end{equation}
where ${\rm k} = 0, \pm 1$ represents the spatial curvature, the $\sigma^2$ term represents the anisotropy ({\it cf.} Section~\ref{sec:anisotropy}), and the dots may include matter, radiation, {\it etc}. The ratios of the scalar field energy density to other components, including curvature and anisotropy, scale as
\begin{equation}
\frac{\rho_\phi}{\rho_{{\rm k}, \sigma, \cdots}} = \frac{a^{-3 (1 + w)}}{a^{-2 \sim 6}} \approx a^{-\frac{2}{p}} \propto (-t)^{-2} \propto H^2 .
\end{equation}
Therefore, after the ekpyrotic phase, all other components including the initial curvature and anisotropy are suppressed by a factor
\begin{equation} \label{eq:efolds}
\frac{(H^2)_\text{ek-end}}{(H^2)_\text{ek-beg}} = \frac{X_\text{ek-end}}{X_\text{ek-beg}} = \frac{V_\text{ek-end}}{V_\text{ek-beg}} \equiv e^{2 N} ,
\end{equation}
where $N$ measures the total number of e-folds of modes that exit the horizon during the ekpyrotic phase \cite{Khoury:2003vb}.

To have a nonsingular bounce, $H$ has to pass from negative to zero, requiring a \emph{bouncing phase} during which $\dot{H} > 0$. According to the second Friedmann equation,
\begin{equation}
\dot{H} = - \tfrac{1}{2} ( \rho + P ) = - \tfrac{1}{2} \rho ( 1 + w ) \; ,
\end{equation}
then, the nonsingular bouncing phase must have $w < -1$, a violation of NEC. This is not possible if the kinetic term $P(X)$ is linear in $X$ as in the ekpyrotic phase, because for the Lagrangian (\ref{eq:L}) the above equation becomes (Appendix (\ref{eq:Friedmann2-bgd}))
\begin{equation} \label{eq:Friedmann2}
\dot{H} = - X P_{,X} \; , \quad X = \tfrac{1}{2} \dot{\phi}^2 \; ,
\end{equation}
which implies $\dot{H} = - \tfrac{1}{2} \dot{\phi}^2 \leq 0$ for $P(X) = X$. The idea behind the form of $P(X)$ in Fig.~\ref{fig:kinetic} is that $X$ exits the linear region after the ekpyrotic phase, and enters the ghost condensate region $X < X_\text{c}$ where $P_{,X} < 0$. To be consistent with the ekpyrotic phase, we must have the relation $X_\text{c} < X_\text{ek-beg}$, which implies
\begin{equation}  \label{eq:XcToXend}
\frac{X_\text{c}}{X_\text{ek-end}} < \frac{X_\text{ek-beg}}{X_\text{ek-end}} = e^{-2N} .
\end{equation}
This exponential factor lies at the core of the problems that we will show in this paper.

The transition from the ekpyrotic phase to the bouncing phase is mediated by a brief \emph{kinetic phase} between the two. As in Fig.~\ref{fig:potential}, after the ekpyrotic phase, the potential $V(\phi)$ rises sharply in order to slow down the field $\phi$ and reduce $X$ from $X_\text{ek-end}$ to $X_\text{c}$. Accordingly, the equation of state changes from $w \gg 1$ to $w = -1$. This transient phase lasts much shorter than a Hubble time, during which the total energy is almost conserved. Let $V_\text{c}$ be the value of the potential when $X$ reaches $X_\text{c}$, then we have
\begin{equation}
V_\text{c} = 3 H_\text{c}^2 \approx 3 H_\text{ek-end}^2 \approx 3 p \, |V_\text{ek-end}| \, ,
\end{equation}
which means $V_\text{c}$ is smaller but of a similar order of magnitude to $|V_\text{ek-end}|$.

\begin{figure}
\centering
\includegraphics[width=0.4\textwidth]{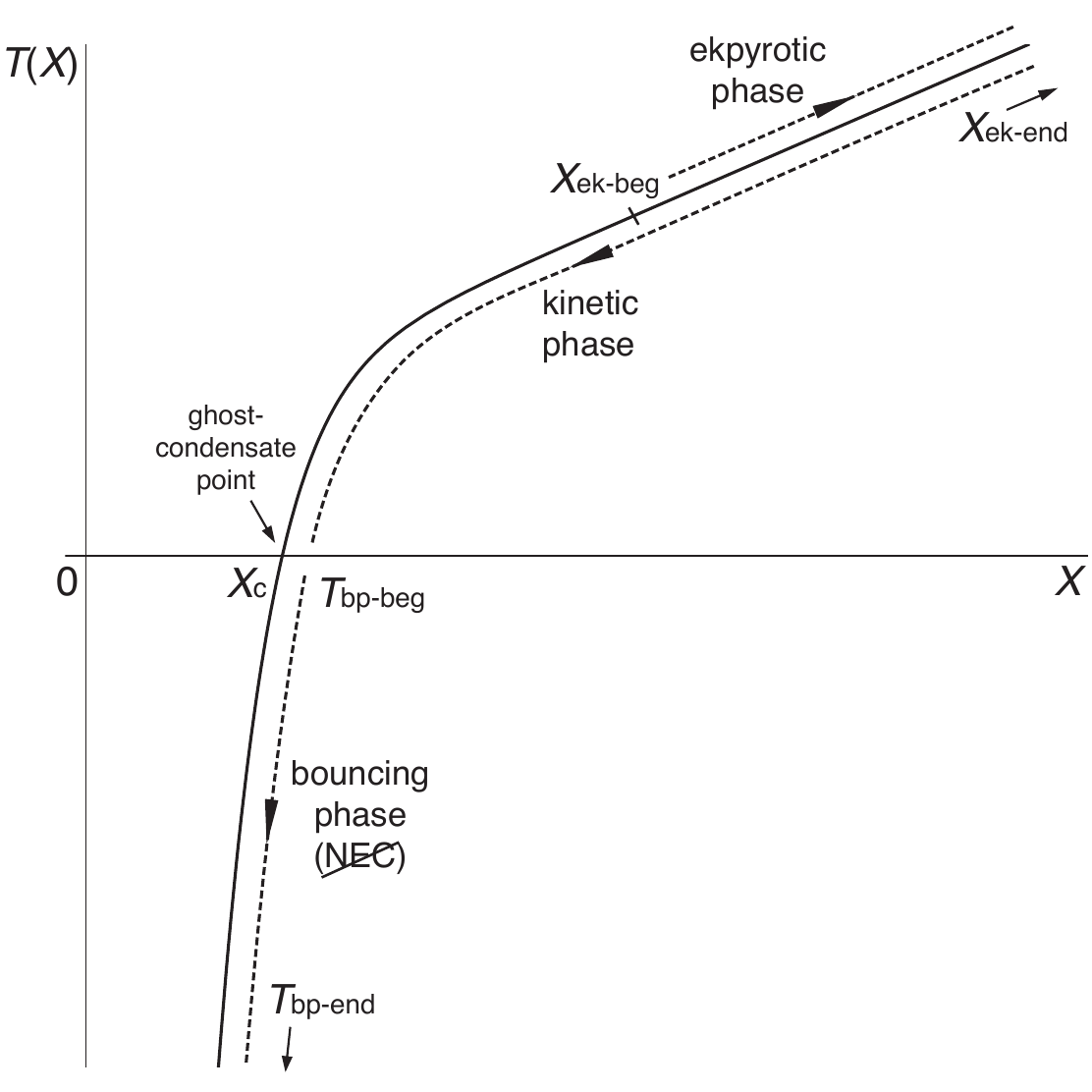}
\caption{The kinetic energy $T(X) = 2 X P_{,X} - P$ for the scalar field $\phi$. It is a linear function of $X$ in the region $X \gg X_\text{c}$, but becomes negative at $X < X_\text{c}$.}
\label{fig:density}
\end{figure}

The \emph{bouncing phase} begins when $X$ further decreases to less than $X_\text{c}$, whereby the equation of state $w$ falls below $-1$. The kinetic energy of the scalar field,
\begin{equation}
T(X) = 2X P_{,X} - P \ ,
\end{equation}
becomes negative, as shown in Fig.~\ref{fig:density}.
The bounce occurs when $X$ decreases to the point where the negative kinetic energy $T$ cancels the positive potential energy $V$, giving $H^2 = \tfrac{1}{3} (T + V) = 0$. Here the negative kinetic energy does not incur ghost instability around $X \approx X_\text{c}$ as long as $T_{,X} = P_{,X} + 2 X P_{,XX}$ is positive \cite{ArkaniHamed:2003uy}. This is satisfied if $P(X)$ grows quickly as $X$ decreases; in particular, if
\begin{equation} \label{eq:PXX}
X P_{,XX} \gg - P_{,X} \; , \quad \hbox{for } X < X_\text{c} \; ,
\end{equation}
which implies a large slope in $T$. In that case the bounce would happen at an $X$ very close to $X_\text{c}$ where the ghost condensation model is valid.

We note that the speed of sound,
\begin{equation} \label{eq:cs2}
c_s^2 = \frac{P_{,X}}{T_{,X}} = \frac{P_{,X}}{2X P_{,XX} + P_{,X}} \; ,
\end{equation}
takes the canonical value $1$ in the ekpyrotic phase when $P(X)$ is linear, but becomes negative in the bouncing phase when $X < X_\text{c}$. Condition (\ref{eq:PXX}) is equivalent to $| c_s^2 | \ll 1$ for $X < X_\text{c}$, which limits the rate at which gravitational instability develops when NEC is violated. In our analysis, we will avoid the gravitational instability by assuming an extremely small value of $| c_s^2 |$ in the bouncing phase, as quantified in Section~\ref{sec:curvature}. Then we can focus on problems of growing curvature and anisotropy that arise regardless of the gravitational instability.

\section{The bouncing solution} \label{sec:bounce}

Before studying the evolution of perturbations, a more detailed description of the bouncing phase is needed. One major issue is the curvature and anisotropy terms in the Friedmann equation (\ref{eq:Friedmann1}). The curvature and anisotropy are exponentially suppressed during the ekpyrotic phase, but they start to grow faster than the scalar field energy when $w < -1$ during the bouncing phase. It is commonly assumed that the bouncing phase can be made as short as a few Hubble times, so that the curvature and anisotropy remain negligible. However, the bouncing phase cannot be made arbitrarily short. By the end of the ekpyrotic phase, the Hubble parameter $H$ is negative and exponentially large, so there must be a long period with $\dot{H} > 0$ in order for $H$ to increase to zero.

To determine how long the bouncing phase really lasts, we solve the equation of motion for the scalar field,
\begin{equation} \label{eq:EOMX}
T_{,X} \, \dot{X} + 6 H P_{,X} \, X + V_{,\phi} \, \dot{\phi} = 0 .
\end{equation}
Using (\ref{eq:PXX}), a general solution to Eq.~(\ref{eq:EOMX}) can be found that describes a nonsingular bounce. First notice that Eq.~(\ref{eq:PXX}) implies, upon integration from $X_\text{c}$,
\begin{equation} \label{eq:PX}
| X P_{,X} | \gg P \; , \quad \hbox{for } X < X_\text{c} \; .
\end{equation}
Therefore the kinetic energy becomes
\begin{equation} \label{eq:approxT}
T(X) = 2 X P_{,X} - P \approx 2 X P_{,X} \; .
\end{equation}
The potential energy and its gradient are nearly constant during the bouncing phase,
\begin{equation} \label{eq:approxV}
V(\phi) \approx V_\text{c} \; , \quad V_{,\phi} \approx V_{,\phi_\text{c}} \; ,
\end{equation}
since
\begin{equation}
\frac{\Delta V}{V_\text{c}} \approx \Big( \frac{- V_{,\phi_\text{c}}}{V_\text{c}} \Big) \Delta \phi \approx \Big( \frac{- V_{,\phi_\text{c}}}{V_\text{c}} \Big) \sqrt{2 X_\text{c}} \, \Delta t_\text{bp} \lesssim N \, e^{-N} ,
\end{equation}
and similarly for $V_{,\phi}$, where $\Delta t_\text{bp}$ is the duration of the bouncing phase as found below. The factors $\big| \frac{- V_{,\phi_\text{c}}}{V_\text{c}} \big|$ and $\big| \frac{V_{,\phi_\text{c} \phi_\text{c}}}{V_\text{c}} \big|$ are taken to be much greater than $1$ yet much less than $e^N$; otherwise one has to fine-tune the steepness of the potential to super Planckian scales, which is unphysical as we shall argue in the end of the paper.

We are now in a position to solve the equation
\begin{equation} \label{eq:EOM}
T_{,X} \, \dot{X} + 3 H T + V_{,\phi_\text{c}} \, \dot{\phi} = 0 .
\end{equation}
The solution can be described in three stages according to whether the friction or the gradient term dominates. At the very beginning of the bouncing phase, $|H| \approx |H_\text{c}| = \sqrt{V_\text{c} / 3}$ and $|T|$ is small, hence the friction term is negligible. Next, the negative kinetic energy $|T|$ increases and the friction term overtakes the gradient. Finally, very close to the bounce we have $T \approx - V_\text{c}$ but $|H|$ becomes small, so that the friction term is again subdominant to the gradient term. Among these three stages, the first and the last are both very short and do not contain interesting features. Indeed, since the friction and the gradient terms have the same sign during the bouncing phase, we have
\begin{equation} \label{eq:Tdot}
|\dot{T}| \geq V_{,\phi} \, \dot{\phi} \approx ( - V_{,\phi_\text{c}} ) \sqrt{2 X_\text{c}} \; .
\end{equation}
Denote
\begin{align}
|T_\text{bp-beg}| &\equiv \Big( \frac{- V_{,\phi_\text{c}}}{V_\text{c}} \Big) \sqrt{2 X_\text{c} V_\text{c} / 3} \; , \label{eq:bp-beg} \\
|H_\text{bp-end}| &\equiv \Big( \frac{- V_{,\phi_\text{c}}}{V_\text{c}} \Big) \sqrt{2 X_\text{c}} \; ; \label{eq:bp-end}
\end{align}
then the first stage goes from $T = 0$ until $T \approx T_\text{bp-beg}$, which lasts less than a Hubble time,
\begin{equation}
t_\text{bp-beg} - t_\text{c} \sim \frac{|T_\text{bp-beg}|}{|\dot{T}|} \lesssim \frac{( - V_{,\phi_\text{c}} ) \sqrt{2 X_\text{c} / 3 V_\text{c}}}{( - V_{,\phi_\text{c}} ) \sqrt{2 X_\text{c}}} \approx \frac{1}{3 |H_\text{c}|} \; ,
\end{equation}
hence the scale factor does not change much, $a_\text{bp-beg} \approx a_\text{c} \approx a_\text{ek-end}$. Similarly, the third stage begins from $T \approx T_\text{bp-end} = - V_\text{c} + 3 H_\text{bp-end}^2$, and the time it takes to reach the bounce can be bounded by
\begin{align}
t_\text{b} - t_\text{bp-end} &\sim \frac{3 H_\text{bp-end}^2}{|\dot{T}|} \lesssim \frac{(-V_{,\phi_\text{c}} / V_\text{c})^2 \, 6 X_\text{c}}{( - V_{,\phi_\text{c}} ) \sqrt{2 X_\text{c}}} \nonumber \\
&\sim \Big( \frac{- V_{,\phi_\text{c}}}{V_\text{c}} \Big) \sqrt{\frac{2 X_\text{c}}{p X_\text{ek-end}}} \frac{1}{|H_\text{c}|} \lesssim \frac{e^{-N}}{|H_\text{c}|} \; ,
\end{align}
which is much less than a Hubble time, implying $a_\text{b} \approx a_\text{bp-end}$. Finite factors like $p$ and $\big( \frac{- V_{,\phi_\text{c}}}{V_\text{c}} \big)$ are neglected in these estimates.

Therefore the bouncing phase mainly consists of the middle stage during which the friction term dominates and the gradient is negligible. Under this condition, the equation of motion simplifies to
\begin{equation} \label{eq:EOMT}
\dot{T} + 3 H T = 0 , \quad \hbox{with } H^2 = \tfrac{1}{3} ( T + V_\text{c} ) .
\end{equation}
The bouncing solution is readily given by
\begin{align}
T &= \frac{- V_\text{c}}{\cosh^2 \big( \tfrac{3}{2} |H_\text{c}| (t - t_\text{0}) \big)} \; , \label{eq:bounceT} \\
H &= |H_\text{c}| \, \tanh \big( \tfrac{3}{2} |H_\text{c}| (t - t_\text{0}) \big) \; , \label{eq:bounceH}
\end{align}
where the bounce is formally at $t = t_\text{0}$. This solution applies between the moments $t_\text{bp-beg}$ and $t_\text{bp-end}$, which correspond to $|T| \sim |T_\text{bp-beg}|$ and $|H| \sim |H_\text{bp-end}|$ respectively. From the expression (\ref{eq:bp-beg}) and use Eq.~(\ref{eq:bounceT}), we have
\begin{equation}
\cosh^{-2} \big( \tfrac{3}{2} |H_\text{c}| (t_\text{bp-beg} - t_\text{0}) \big) \sim \Big( \frac{- V_{,\phi_\text{c}}}{V_\text{c}} \Big) \sqrt{\frac{X_\text{c}}{X_\text{ek-end}}} \sim e^{-N} ,
\end{equation}
which gives
\begin{equation}
t_\text{0} - t_\text{bp-beg} \approx \frac{N}{3 |H_\text{c}|} \; .
\end{equation}
Similarly, from the expression (\ref{eq:bp-end}) and use Eq.~(\ref{eq:bounceH}), we find
\begin{equation}
\Big| \tanh \big( \tfrac{3}{2} |H_\text{c}| (t_\text{bp-end} - t_\text{0}) \big) \Big| \sim \Big( \frac{- V_{,\phi_\text{c}}}{V_\text{c}} \Big) \sqrt{\frac{X_\text{c}}{X_\text{ek-end}}} \sim e^{- N} ,
\end{equation}
which gives
\begin{equation} \label{eq:tbp-end}
t_\text{0} - t_\text{bp-end} \approx \frac{2 e^{-N}}{3 |H_\text{c}|} \, .
\end{equation}
Therefore the middle stage, hence the whole bouncing phase, lasts approximately for a period
\begin{equation} \label{eq:duration}
\Delta t_\text{bp} \approx t_\text{bp-end} - t_\text{bp-beg} \approx \frac{N}{3} \frac{1}{|H_\text{c}|} \, .
\end{equation}

This duration eliminates any hope to complete the bounce within just a few Hubble times. As a result, we expect anisotropies to grow significantly during the bouncing phase. From the bouncing solution (\ref{eq:bounceH}), the scale factor $a(t)$ scales as
\begin{equation} \label{eq:bouncea}
a \propto \cosh^{2/3} \big( \tfrac{3}{2} |H_\text{c}| (t - t_\text{0}) \big) \propto |T|^{-1/3} .
\end{equation}
Before the end of the bouncing phase, it contracts by a factor
\begin{equation} \label{eq:abp-end}
\frac{a_\text{bp-end}}{a_\text{bp-beg}} = \Big| \frac{T_\text{bp-end}}{T_\text{bp-beg}} \Big| ^{-\frac{1}{3}} \sim \Big( \frac{- V_{,\phi_\text{c}}}{V_\text{c}} \Big)^{\frac{1}{3}} \Big( \frac{X_\text{c}}{X_\text{ek-end}} \Big) ^{\frac{1}{6}} \lesssim e^{- \frac{1}{3} N} .
\end{equation}
Therefore, naively, the anisotropy term in the Friedmann equation (\ref{eq:Friedmann1}) would increase by a factor
\begin{equation} \label{eq:sigmabpendtobeg}
\frac{(\sigma^2)_\text{bp-end}}{(\sigma^2)_\text{bp-beg}} = \Big( \frac{a_\text{bp-end}}{a_\text{bp-beg}} \Big) ^{-6} \sim \frac{X_\text{ek-end}}{X_\text{c}} \gtrsim e^{2 N} ,
\end{equation}
which completely cancels the suppression (\ref{eq:efolds}) it has experienced during the ekpyrotic phase. That is, the anisotropy has returned!

To be more precise, during the ekpyrotic phase the suppression of anisotropy happens in such a way that the anisotropy term itself remains almost constant, while the scalar field energy increases by an exponential factor $e^{2N}$. In the bouncing phase, however, the anisotropy term grows by the same factor $e^{2N}$, while the scalar field energy decreases rapidly towards $0$. Thus there must be a point near the end of the bouncing phase where the anisotropy contribution to the Friedmann equation (\ref{eq:Friedmann1}) overtakes the scalar field energy. After this point, the nonsingular bouncing solution (\ref{eq:bounceH}) that assumes the dominance of the scalar field becomes invalid. To see how the exponential growth of anisotropy may disrupt the nonsingular bounce, we need to study perturbations around the background evolution found above.

\section{Curvature perturbation and power spectrum} \label{sec:curvature}

Consider linear perturbations about the metric (\ref{eq:FRW}),
\begin{align} \label{eq:pFRW}
ds^2 = &\ a(\tau)^2 \Big[ - ( 1 + 2 A) d\tau^2 + 2 ( B_{,i} + S_i ) d\tau dx^i \\
& + \big( ( 1 - 2 \psi ) \delta_{ij} + 2 E_{,ij} + 2 F_{(i,j)} + 2 h_{ij} \big) dx^i dx^j \Big] , \nonumber
\end{align}
where $\tau$ is the conformal time, $dt = a \, d\tau$. We will denote conformal time derivative as $^\prime \equiv \frac{d}{d \tau} = a \frac{d}{dt}$, and spatial Laplacian as $\nabla^2 \equiv \partial^i \partial_i$. The functions $A$, $B$, $\psi$ and $E$ represent the scalar perturbations; $S_i$ and $F_i$, with $S^i_{,i} = F^i_{,i} = 0$, represent the vector perturbations; and $h_{ij}$, with $h^i_{\, i} = h^i_{\, j,i} = 0$, represent the tensor perturbations. The definition of these perturbative quantities are subject to gauge transformations of the coordinates, as discussed in Appendix~\ref{sec:perturbation}. The constant time hypersurface is given by the normal $n_\mu = (-a(1+A), \vec{0}\,)$. The perturbation of its intrinsic curvature is (Appendix (\ref{eq:3R_pert}))
\begin{equation} \label{eq:intrinsic-curv}
\delta ^{(3)} \! R = \frac{4}{a^2} \nabla^2 \psi \ ,
\end{equation}
hence $\psi$ is referred to as the curvature perturbation. The shear of the hypersurface is given by (Appendix (\ref{eq:sigma_pert}))
\begin{equation}
\sigma_{ij} = a \big[ ( \sigma^\text{S}_{,ij} - \tfrac{1}{3} \delta_{ij} \nabla^2 \sigma^\text{S} ) + \sigma^\text{V}_{(i,j)} \big] ,
\end{equation}
where the scalar and vector contributions are
\begin{align}
\sigma^\text{S} &= E' - B \ , \label{eq:shear-s} \\
\sigma^\text{V}_i &= F_i ' - S_i \ , \label{eq:shear-v}
\end{align}
which will be related to the anisotropy. The evolution of these perturbative quantities is determined by the Einstein equation and the equation of motion for the scalar field $\phi$ and its perturbation $\delta \phi$, as shown in Appendix~\ref{sec:perturbation}.

The evolution of the curvature perturbation involves only the scalar perturbations. To agree with observations, the contracting phase of the Universe has to generate a nearly scale-invariant spectrum of curvature perturbations. Several mechanisms have been proposed that can give rise to such scale-invariant curvature perturbation, including the multi-field entropic mechanism \cite{Lehners:2007ac, Buchbinder:2007ad} and the single-field adiabatic ekpyrotic mechanism \cite{Khoury:2009my, Khoury:2011ii}. Once the generated scale-invariant mode exits the horizon, it is expected to be conserved on superhorizon scales. Meanwhile, there exists an adiabatic contribution to the curvature perturbation that is time-varying according to the equation of state. This adiabatic contribution generally has a blue spectrum, and is subdominant to the constant scale-invariant modes on large scales. However, a serious problem arises when the equation of state $w$ abruptly changes in the transition from the ekpyrotic phase to the bouncing phase. It happens that an adiabatic mode grows exponentially as $w$ drops past $-1$, and surpasses the scale-invariant mode even on large scales.

To observe this problem, consider the curvature perturbation in the comoving gauge defined by $\delta \phi = 0$, as is done in \cite{Xue:2010ux}. The curvature perturbation in this gauge is described by the gauge-invariant quantity (Appendix (\ref{eq:psicom}))
\begin{equation} \label{eq:R}
\mathcal{R} \equiv \psi + H \frac{\delta \phi}{\dot{\phi}} \; .
\end{equation}
The Fourier mode $\mathcal{R}_k$ with comoving wavenumber $k$ obeys the equation (Appendix (\ref{eq:Rcom}))
\begin{equation} \label{eq:Rk}
\mathcal{R}_k^{\prime \prime} + 2 \frac{z^\prime}{z} \mathcal{R}_k^{\prime} + c_s^2 k^2 \mathcal{R}_k = 0 ,
\end{equation}
where $z = a \sqrt{{-2 \dot{H}}/{c_s^2 H^2}}$. After the mode exits the horizon, the $k^2$ term can be treated perturbatively, and the equation is formally solved by expanding in orders of $k^2$,
\begin{equation} \label{eq:Rexpansion}
\mathcal{R}_k = \mathcal{R}_k^{(0)} - k^2 \int \frac{d \tau}{z^2} \int d\tau \, c_s^2 z^2 \mathcal{R}_k \; .
\end{equation}
The leading order $\mathcal{R}_k^{(0)}$ is the solution to the equation without the $k^2$ term, which contains two general solutions,
\begin{equation} \label{eq:R0}
\mathcal{R}_k^{(0)} = C_1 (k) + C_2 (k) \int \frac{d \tau}{z^2} \equiv \mathcal{R}_k^\text{const} + \mathcal{R}_k^\text{int} \, .
\end{equation}
These two terms are the leading adiabatic contributions to the curvature perturbation; the $k$-dependence of the dominant term determines the spectral index of adiabatic perturbations on large scales.

The $k$-dependence of the constants $C_1, C_2$ can be found by matching to initial conditions in Minkowski vacuum when the mode originated from quantum fluctuations deep inside the horizon. Introducing the canonical variable $v_k \equiv z \, \mathcal{R}_k$, Eq.~(\ref{eq:Rk}) becomes
\begin{equation} \label{eq:v}
v_k^{\prime \prime} + \big( c_s^2 k^2 - \frac{z''}{z} \big) v_k = 0 .
\end{equation}
During the ekpyrotic phase, $c_s^2 = 1$ and the scaling solution (\ref{eq:scaling}) gives $z \sim (-\tau)^\frac{p}{1-p}$, so the \emph{freeze-out horizon} scale is given by
\begin{equation} \label{eq:zppz-ek}
\Big| \frac{z''}{z} \Big| = \frac{p \, (1-2p)}{(1-p)^2 \, \tau^2} \, .
\end{equation}
Then the solution to Eq.~({\ref{eq:v}}) is given by Hankel functions,
\begin{equation}
v_k (\tau) = \sqrt{x} \big[ A H_\nu^{(1)} (x) + B H_\nu^{(2)} (x) \big] ,
\end{equation}
where $x \equiv k (-\tau)$, and $\nu = \tfrac{1}{2} - \tfrac{p}{1-p}$. At early times when the mode is deep inside the horizon, $x \gg 1$, the solution should be matched to the vacuum solution for a Minkowski background \cite{Birrell:1982ix},
\begin{equation}
v_k \rightarrow \frac{1}{\sqrt{2k}} e^{-i k \tau} , \quad |k \tau| \rightarrow \infty .
\end{equation}
Using the asymptotic behavior of Hankel functions,
\begin{equation}
H_\nu^{(1,2)} (x) \rightarrow \sqrt{\frac{2}{\pi x}} \, e^{\pm i ( x - \frac{\nu \pi}{2} - \frac{\pi}{4})} , \quad x \rightarrow \infty,
\end{equation}
we can fix the constants $B = 0$, $A = \sqrt{\frac{\pi}{4k}} \, e^{i ( \frac{\nu \pi}{2} + \frac{\pi}{4})}$. Therefore, neglecting the constant phase factor, we have
\begin{equation}
v_k (\tau) = \sqrt{\frac{\pi x}{4 k}} \, H_\nu^{(1)} (x) .
\end{equation}
After the mode exits the horizon, the solution approaches the other limit, $x \ll 1$, where it allows the expansion
\begin{align}
v_k = \sqrt{\frac{\pi x}{4 k}} \bigg[ & -i \frac{2^\nu \Gamma(\nu)}{\pi x^\nu} + \Big( - i \frac{\cos(\nu \pi) \Gamma(-\nu)}{2^\nu \pi} \\
& + \frac{1}{2^\nu \Gamma(1+\nu)} \Big) x^\nu + O(x^{2-\nu}) \bigg] , \quad x \rightarrow 0 . \nonumber
\end{align}
From this, the expression for $\mathcal{R}_k$ becomes
\begin{align}
\mathcal{R}_k \approx & \, \sqrt{\frac{\pi p}{8}} \frac{(-\tau)^{\frac{1}{2} - \frac{p}{1-p}}}{(1-p)^{\frac{p}{1-p}} \alpha^{\frac{1}{1-p}}} \bigg[ -i \frac{2^\nu \Gamma(\nu)}{\pi} k^{-\nu} (-\tau)^{-\nu} \nonumber \\
& + \Big( i \frac{\cos(\nu \pi) \Gamma(1-\nu)}{2^\nu \pi \nu} + \frac{1}{2^\nu \nu \Gamma(\nu)} \Big) k^{\nu} (-\tau)^{\nu} \bigg] \nonumber \\
\approx & \, \frac{-i \sqrt{p}}{2 \alpha^{\frac{1}{1-p}}} k^{-\frac{1}{2} + \frac{p}{1-p}} + \frac{\sqrt{p}}{2 \alpha^{\frac{1}{1-p}}} k^{\frac{1}{2} - \frac{p}{1-p}} (-\tau)^{\frac{1-3p}{1-p}},
\end{align}
in the limit of small $p$, but keeping the exact powers. These leading terms correspond to the solution (\ref{eq:R0}),
\begin{align} \label{eq:Rconst+Rint}
\mathcal{R}_k^\text{const} + \mathcal{R}_k^\text{int} &= C_1 (k) + C_2 (k) \int_0^\tau \frac{c_s^2 H^2}{a^2 (-2 \dot{H})} d\tau \\
&= C_1 (k) + C_2 (k) \frac{p \, \alpha^{\frac{-2}{1-p}}}{2-6p} \big( (1-p)(-\tau) \big)^{\frac{1-3p}{1-p}} . \nonumber
\end{align}
Comparing the coefficients, one obtains
\begin{align}
C_1 (k) &\approx \frac{-i \sqrt{p}}{2 \alpha^{\frac{1}{1-p}}} \, k^{-\frac{1}{2} + \frac{p}{1-p}} \sim \frac{1}{\sqrt{k}} \; , \\
C_2 (k) &\approx \frac{\alpha^{\frac{1}{1-p}}}{\sqrt{p}} \, k^{\frac{1}{2} - \frac{p}{1-p}} \sim \sqrt{k} \; . \label{eq:C2}
\end{align}
Since $k^{3/2} |C_1 (k)| \sim k$, and $k^{3/2} |C_2 (k)| \sim k^2$, both $\mathcal{R}_k^\text{const}$ and $\mathcal{R}_k^\text{int}$ terms have blue spectral indices.

To produce a scale-invariant contribution to the curvature perturbation, we may, for instance, invoke the entropic mechanism described in \cite{Buchbinder:2007ad}. In this mechanism, an independent scale-invariant contribution $\mathcal{R}_k^\text{sc-inv}$ is generated from entropic perturbations by an additional scalar field at the end of the ekpyrotic phase, where
\begin{equation}
\mathcal{R}_k^\text{sc-inv} \sim \frac{1}{k^{3/2}} \, .
\end{equation}
Thus the total curvature perturbation is the sum of the terms,
\begin{equation}
\mathcal{R}_k^\text{tot} \approx \mathcal{R}_k^\text{sc-inv} + \mathcal{R}_k^\text{const} + \mathcal{R}_k^\text{int} .
\end{equation}

By the end of the ekpyrotic phase, according to Eq.~(\ref{eq:Rconst+Rint}),
\begin{align}
& \mathcal{R}_k^\text{const} \big|_\text{ek-end} = C_1 (k) , \\
& \mathcal{R}_k^\text{int} \big|_\text{ek-end} \approx \frac{C_2 (k) \, p^2}{2 (a^3 H)_{\text{ek-end}}} \approx \frac{C_2 (k)}{2 a_\text{ek-end}^3} \sqrt{\frac{p^3}{X_\text{ek-end}}} \; . \label{eq:ek-end}
\end{align}
Hence $\mathcal{R}_k^\text{int}$ is suppressed relative to $\mathcal{R}_k^\text{const}$ by a factor
\begin{equation} \label{eq:Nk}
\bigg| \frac{\mathcal{R}_k^\text{int}}{\mathcal{R}_k^\text{const}} \bigg|_\text{ek-end} \approx \bigg| \frac{p \, k}{(aH)_{\text{ek-end}}} \bigg|
\approx \sqrt{\frac{X_k}{X_{\text{ek-end}}}} \equiv  \, e^{-N_k} ,
\end{equation}
where $X_k$ is the kinetic energy at horizon crossing and $N_k$ is the remaining number of e-folds of the ekpyrotic phase after the $k$-mode exits the horizon. Due to its blue spectrum, the $\mathcal{R}_k^\text{const}$ term is in turn dominated by the scale-invariant term,
\begin{equation}
\bigg| \frac{\mathcal{R}_k^\text{const}}{\mathcal{R}_k^\text{sc-inv}} \bigg|_\text{ek-end} \sim \bigg| \frac{k}{(aH)_{\text{ek-end}}} \bigg| \approx \sqrt{ \frac{X_k}{X_{\rm ek-end}} } = e^{-N_k} .
\end{equation}
Hence the integral term is \emph{sub-subdominant},
\begin{equation} \label{eq:REntro}
\bigg| \frac{\mathcal{R}_k^\text{int}}{\mathcal{R}_k^\text{sc-inv}} \bigg|_\text{ek-end} \sim \bigg| \frac{k}{(aH)_{\text{ek-end}}} \bigg|^2 \approx \frac{X_k}{X_{\rm ek-end}} = e^{-2 N_k} .
\end{equation}
Therefore, the total curvature perturbation right after the ekpyrotic phase is
\begin{equation}
\mathcal{R}_k^\text{tot} \Big|_\text{ek-end} \approx \mathcal{R}_k^\text{sc-inv} \sim \frac{1}{k^{3/2}} \, ,
\end{equation}
which confirms that the entropic mechanism produces a scale-invariant spectrum. We note that the $\mathcal{R}_k^\text{sc-inv}$ and $\mathcal{R}_k^\text{const}$ terms remain constant afterwards, but the integral term $\mathcal{R}_k^\text{int}$ is time-varying even on large scales. The total curvature perturbation $\mathcal{R}_k$ is conserved and scale-invariant on superhorizon scales only if the time-varying piece remains negligible.

We now demonstrate that, in fact, the $\mathcal{R}_k^\text{int}$ term grows rapidly in the kinetic phase when $\dot{H}$ increases from negative to $0$ and $w$ crosses $-1$. Indeed, from Eqs.~(\ref{eq:Friedmann2}) and (\ref{eq:cs2}), the integral term can be written as
\begin{equation} \label{eq:Rint}
\mathcal{R}_k^\text{int} = C_2(k) \int \frac{c_s^2}{a^3} \frac{H^2}{-2\dot{H}} dt = C_2 (k) \int \frac{1}{a^3} \frac{H^2}{2X T_{,X}} \frac{dX}{\dot{X}} \, .
\end{equation}
As the field passes the bottom of the potential and climbs up the other side, its kinetic energy quickly decreases to $X \approx X_\text{c}$ where $P(X)$ becomes nonlinear. During this rapid phase the friction term in Eq.~(\ref{eq:EOMX}) can be neglected, hence
\begin{align} \label{eq:Rw-1}
\mathcal{R}_k^\text{int} \Big|_{w \rightarrow -1} &\approx C_2 (k) \int \frac{1}{a^3} \frac{H^2}{2X} \frac{dX}{V_{,\phi} \sqrt{2X}} \nonumber \\
&\approx C_2 (k) \frac{1}{2 a^3_\text{c}} \Big( \frac{H^2_\text{c}}{V_{,\phi_\text{c}}} \Big) \int^{X_\text{c}} \frac{dX}{\sqrt{2 X^3}} \nonumber \\
&\approx \frac{C_2 (k)}{3 a^3_\text{ek-end}} \Big( \frac{V_\text{c}}{- V_{,\phi_\text{c}}} \Big) \frac{1}{\sqrt{2 X_\text{c}}} \; ,
\end{align}
where the integral in the first line is dominated by contributions from near the upper limit, and so we approximated the slowly varying quantities by their values there. Compared to the value in (\ref{eq:ek-end}) at the end of the ekpyrotic phase, the integral term has now been exponentially amplified,
\begin{equation} \label{eq:Rw-1toEnd}
\frac{|\mathcal{R}_k^\text{int}|_{w \rightarrow -1}}{|\mathcal{R}_k^\text{int}|_{\text{ek-end}}} \approx \Big( \frac{V_\text{c}}{- V_{,\phi_\text{c}}} \Big) \sqrt{\frac{X_\text{ek-end}}{X_\text{c}}} \gtrsim e^N .
\end{equation}
It exceeds the scale-invariant term by a ratio
\begin{equation} \label{eq:Rw-1toEntro}
\left| \frac{\mathcal{R}_k^\text{int}}{\mathcal{R}_k^\text{sc-inv}} \right| _{w \rightarrow -1} \sim \frac{X_k}{X_\text{ek-end}} \, \sqrt{\frac{X_\text{ek-end}}{X_\text{c}}} \gtrsim e^{N - 2 N_k} > 1 ,
\end{equation}
for a wide range of modes with $N_k < N/2$.

Thus, modes that exit the horizon in the second half of the ekpyrotic phase, including all modes within current observable horizon, are overwhelmed by the integral term which carries a blue spectrum,
\begin{equation}
\mathcal{R}_k^\text{tot} \Big|_{w \rightarrow -1} \approx \mathcal{R}_k^\text{int} \sim \sqrt{k} \; ,
\end{equation}
in contradiction to observations. Moreover, assuming that the amplitude of the scale-invariant modes in (\ref{eq:Rw-1toEntro}) matches the amplitude measured by the cosmic microwave background (CMB), $\Delta_{\mathcal{R}^\text{sc-inv}}^2 = \frac{k^3}{2\pi^2} |\mathcal{R}_k^\text{sc-inv}|^2 \approx 2.2~\times~10^{-9}$ \cite{Lehners:2007ac}, then the blue modes will have an amplitude $\Delta_\mathcal{R}^2 \approx e^{2 N - 4 N_k} \Delta_{\mathcal{R}^\text{sc-inv}}^2$, which is already nonlinear for modes with $N_k < N/2 - 5$. Note that these modes remain outside the horizon after the ekpyrotic phase, since the horizon continues to shrink in the kinetic phase. This can be seen by writing
\begin{align} \label{eq:zppz}
\Big| \frac{z''}{z} \Big| = a^2 H^2 \Big[ &\ 3 \Big( \frac{V_{,\phi\phi}}{3 H^2} \Big) - 6 \sqrt{3(1+w)} \Big( \frac{- V_{,\phi}}{3 H^2} \Big) \nonumber \\
& - 2 + \frac{21}{2} (1+w) - \frac{9}{2} (1+w)^2 \Big] ,
\end{align}
which is dominated by the first term as $w \rightarrow -1$,
\begin{equation}
\Big| \frac{z''}{z} \Big| \approx 3 \Big( \frac{V_{,\phi_\text{c}\phi_\text{c}}}{V_\text{c}} \Big) (aH)^2_\text{ek-end} .
\end{equation}
Since we assumed a steep potential in the kinetic phase, this expression is generally larger than the value at the end of the ekpyrotic phase, $| z''/z | \approx (aH)^2_\text{ek-end} / p$, from Eq.~(\ref{eq:zppz-ek}).

It remains to show that the dominantly blue curvature perturbation is maintained through the bouncing phase without significant changes. By Eq.~(\ref{eq:Rint}), the change in the integral term is
\begin{equation} \label{eq:Rint2}
\Delta \mathcal{R}_k^\text{int} = C_2 (k) \int \frac{H^2}{2 a^3 \dot{T}} \frac{dX}{X} \, .
\end{equation}
We observe that during the whole bouncing phase we have the relation
\begin{equation}
\frac{H}{a^3 \dot{T}} \leq \text{const} \approx \frac{1}{3 a_\text{ek-end}^3} \Big( \frac{V_\text{c}}{-V_{,\phi_\text{c}}} \Big) \frac{1}{|H_\text{c}| \sqrt{2 X_\text{c}}} \, .
\end{equation}
The equality holds for the bouncing solution due to Eq.~(\ref{eq:EOMT}) and the scaling relation (\ref{eq:bouncea}); the value of the constant is found at the beginning of the bouncing phase from Eq.~(\ref{eq:bp-beg}). The inequality is true in the short stages before $t_\text{bp-beg}$ and after $t_\text{bp-end}$ as a result of Eq.~(\ref{eq:Tdot}). Therefore we can estimate Eq.~(\ref{eq:Rint2}) as
\begin{align} \label{eq:Rint-bp}
\Delta \mathcal{R}_k^\text{int} &\lesssim \frac{C_2 (k)}{6 a_\text{ek-end}^3} \Big( \frac{V_\text{c}}{-V_{,\phi_\text{c}}} \Big) \frac{1}{\sqrt{2 X_\text{c}}} \int \frac{H}{|H_\text{c}|} \frac{dX}{X} \nonumber \\
&\lesssim \frac{C_2 (k)}{3 a_\text{ek-end}^3} \Big( \frac{V_\text{c}}{-V_{,\phi_\text{c}}} \Big) \frac{1}{\sqrt{2 X_\text{c}}} \bigg| \frac{\Delta X}{2 X_\text{c}} \bigg| \; ,
\end{align}
where we used the fact that $|H| \leq |H_\text{c}|$ in the bouncing phase, and assumed that $\Delta X \ll X_\text{c}$ in order for the ghost condensate model to be valid. Hence the total change of the integral term is much less than the value (\ref{eq:Rw-1}) obtained in the kinetic phase right before the bouncing phase begins.

Our analysis differs in an important way from \cite{Creminelli:2007aq, Lin:2010pf}, which considered similar models and concluded that the comoving curvature perturbation changes negligibly near the bounce. In those cases, the bounce is analyzed in the limit that the Hubble parameter varies approximately linearly with time. From our bouncing solution (\ref{eq:bounceH}), this corresponds to the period when $|t - t_\text{0}| \ll 1 / |H_\text{c}|$, so that
\begin{equation}
H \approx \tfrac{3}{2} H_\text{c}^2 (t - t_\text{0}) , \quad \dot{H} \approx \tfrac{3}{2} H_\text{c}^2 .
\end{equation}
This is within the last e-fold of the bouncing phase, where $|c_s^2| \ll 1$ and $a \approx a_\text{bp-end}$ given in Eq.~(\ref{eq:abp-end}). During this linear regime the curvature perturbation changes by
\begin{align}
\Delta \mathcal{R}_k^\text{int} &\approx C_2 (k) \int \frac{c_s^2}{a_\text{bp-end}^3} \frac{H^2}{-2\dot{H}} dt \label{eq:R-laste} \\
&\ll \frac{C_2 (k)}{a_\text{bp-end}^3} \int \frac{3}{4} H_\text{c}^2 (t - t_\text{0})^2 dt \nonumber \\
&\ll \frac{C_2 (k)}{a_\text{ek-end}^3} \Big( \frac{V_\text{c}}{-V_{,\phi_\text{c}}} \Big) \sqrt{\frac{X_\text{ek-end}}{X_\text{c}}} \, \frac{1}{4 |H_\text{c}|} \nonumber \\
&\sim \frac{C_2 (k)}{a_\text{ek-end}^3} \Big( \frac{V_\text{c}}{-V_{,\phi_\text{c}}} \Big) \frac{1}{\sqrt{2 X_\text{c}}} \, . \nonumber
\end{align}
Thus, as found by \cite{Creminelli:2007aq, Lin:2010pf}, $\mathcal{R}$ changes very little from its value (\ref{eq:Rw-1}) at the beginning of the bouncing phase. Indeed, we can simply see from the expression (\ref{eq:R-laste}) that the integral term is decaying as $|H| \rightarrow 0$ at this last stage of the bouncing phase. However, we emphasize that the comoving curvature perturbation grows large at a much earlier stage just before the bouncing phase begins. This period was ignored in those previous studies, and so the problem with the blue spectrum was missed.

Finally, let us take care of the gravitational instability in the bouncing phase when $c_s^2 < 0$. During the bouncing phase, perturbation modes may reenter the horizon. For example, in the case of a slowly varying $c_s^2$ ({\it i.e.} $\dot{c_s} / H c_s \ll 1$) consistent with Eq.~(\ref{eq:PXX}), the freeze-out horizon scale corresponding to the bouncing solution (\ref{eq:bounceH}) and (\ref{eq:bouncea}) is given by
\begin{align}
\frac{z''}{z} \approx a^2 H_\text{c}^2 \bigg[ &- \frac{1}{4} - \frac{1}{2 \cosh^2 \big( \tfrac{3}{2} |H_\text{c}| (t - t_\text{0}) \big)} \nonumber \\
&+ \frac{9}{2 \sinh^2 \big( \tfrac{3}{2} |H_\text{c}| (t - t_\text{0}) \big)} \bigg] ,
\end{align}
which goes from negative to positive and crosses zero at $t_0 - t \approx 1.4 / |H_\text{c}|$. Therefore all modes briefly reenter the horizon around this time, and exit again when $t - t_0 \rightarrow 0$ approaching the bounce. According to Eq.~(\ref{eq:v}), modes inside the horizon grow unstable instead of undergoing oscillations. Since the growth rate is proportional to $|c_s| k$, this instability could be tamed if $|c_s^2|$ is small and the duration inside the horizon is short. The dangerous modes are those on small scales, which reenter the horizon early in the bouncing phase and stay inside until very close to the bounce. Their duration inside the horizon can approach the upper bound (\ref{eq:duration}), or in conformal time,
\begin{equation}
|c_s| k \Delta \tau \sim |c_s| \, e^{\frac{N}{3} - N_k} ,
\end{equation}
which may create a problem for the modes with $N_k < N/3$. To avoid this gravitational instability, the speed of sound has to be
\begin{equation} \label{eq:cs2cond}
|c_s^2| \lesssim e^{- \frac{2}{3} N} .
\end{equation}
This is the same condition under which the leading term (\ref{eq:R0}) dominates the expansion in Eq.~(\ref{eq:Rexpansion}) and our computations follow. We note that even when gravitational instability is suppressed under the condition (\ref{eq:cs2cond}), the problem with the exponential growth of the comoving curvature perturbation still occurs just before $c_s^2$ becomes negative.

Having traced the evolution of the comoving curvature perturbation through each stage of the contracting phase, let us now comment on the curvature perturbation in other gauges. Under a general coordinate transformation $x^\mu \rightarrow x^\mu + \xi^\mu$, the curvature perturbation $\psi$ transforms as (Appendix (\ref{eq:psi_transf}))
\begin{equation}
\psi \rightarrow \psi + a H \xi^0 \; .
\end{equation}
An immediate consequence is that the curvature perturbation is gauge-invariant at a nonsingular bounce where $H = 0$; hence results computed in different gauges must agree as they approach the bounce. Therefore the exponential growth of curvature perturbation that we found in the comoving gauge is truly physical and should exist in other gauges as well. However, although the curvature perturbation in different gauges end up with the same value at the bounce, their patterns of growth can be quite different during the contracting phase before the bounce. For example, in the longitudinal gauge (Appendix~\ref{sec:perturbation}) the Newtonian potential $\Phi$ grows exponentially during the ekpyrotic phase, but undergoes no abrupt change in the transitional kinetic phase, then grows further in the bouncing phase to reach the same value as $\mathcal{R}$ at the bounce. In the above computations, we chose to study the comoving curvature perturbation $\mathcal{R}$ because it will be conserved outside the horizon in the expanding phase including the reheating period, and therefore it is directly related to the density and temperature fluctuations that are observed today. It is also a convenient variable to study because it is almost conserved during both the ekpyrotic and the bouncing phases, whereas the exponential growth can be clearly confined and ascribed to the transition between the two phases, during which the equation of state $w$ changes rapidly.

Since the comoving curvature perturbation $\mathcal{R}$ becomes exponentially large right before the bouncing phase, another natural concern is whether the perturbative computation breaks down at this point. We address this question by noting that the perturbation theory is valid if there exists one particular gauge in which the perturbations of all relevant physical quantities remain small; then the variables in other gauges can be formally defined and related to the quantities in this particular gauge through gauge transformations. In our case, the synchronous gauge plays this role. It can be readily checked that the physical quantities, including the curvature perturbation $\psi_\text{s}$, the shear perturbation $\sigma_\text{s}$ (see Section~\ref{sec:anisotropy}), and the matter perturbations $\delta \rho_\text{s}$, $\delta P_\text{s}$, $(\rho + P) \delta u_\text{s}$, all remain finite during the transitional kinetic phase in this gauge. In particular, the curvature perturbation $\psi_\text{s}$ in the synchronous gauge can be related to $\mathcal{R}$ through (Appendix (\ref{eq:R-to-psi_s}))
\begin{equation}
\psi_\text{s} = \mathcal{R} - H \int_0^t \frac{\dot{\mathcal{R}}}{H} \, dt' \; .
\end{equation}
After the $k$ mode ${\psi_\text{s}}_k$ exits horizon, we may insert the leading terms of $\mathcal{R}_k$ from (\ref{eq:R0}) to find (for brevity we will omit the subscript $k$ from here on)
\begin{equation} \label{eq:psis}
\psi_\text{s} \approx C_1 (k) + C_2 (k) \int_0^t dt' \dot{H} \int_0^{t'} dt'' \frac{c_s^2 H}{2 a^3 \dot{H}} \, .
\end{equation}
Due to the presence of $\dot{H}$ in the outer integral, the second term remains small when $\dot{H} \rightarrow 0$, so that the curvature perturbation is well behaved in this gauge when the equation of state $w$ crosses $-1$. Hence it is indeed legitimate to carry our computations beyond this point into the bouncing phase.

Nevertheless, even the synchronous curvature perturbation $\psi_\text{s}$ has to grow exponentially near the bounce to match the value of $\mathcal{R}$. To prove this, let us evaluate Eq.~(\ref{eq:psis}) in the bouncing phase as follows. First denote
\begin{equation} \label{eq:psis-const+int}
\psi_\text{s}^\text{const} + \psi_\text{s}^\text{int} \equiv C_1 (k) + C_2 (k) \int_0^t dt' \dot{H} \cdot I(t') \, ,
\end{equation}
where $I(t')$ is the inner integral in (\ref{eq:psis}). In the kinetic phase, similar to Eq.~(\ref{eq:Rw-1}), $I(t')$ can be computed as
\begin{align}
I(t') &\equiv \int_0^{t'} dt'' \frac{c_s^2 H}{2 a^3 \dot{H}} \approx \frac{H_\text{c}}{2 a_\text{c}^3} \int \frac{dX / X}{(-V_{,\phi}) \sqrt{2X}} \nonumber \\
&\xrightarrow{t' \rightarrow t_\text{c}} \frac{1}{3 a^3_\text{ek-end}} \Big( \frac{V_\text{c}}{- V_{,\phi_\text{c}}} \Big) \frac{1}{|H_\text{c}| \sqrt{2 X_\text{c}}} \, .
\end{align}
Here this huge growth as $t' \rightarrow t_\text{c}$ is tempered by $\dot{H} \rightarrow 0$ in the outer integral in (\ref{eq:psis-const+int}), hence does not show in $\psi_\text{s}^\text{int}$. Then in the bouncing phase, as for Eqs.~(\ref{eq:Rint2}) and (\ref{eq:Rint-bp}), $I(t')$ can be found to stay nearly constant,
\begin{align}
\Delta I &\approx \int_{X_\text{c}}^X \frac{H}{2 a^3 \dot{T}} \frac{dX}{X} \nonumber \\
&\lesssim \frac{1}{3 a_\text{ek-end}^3} \Big( \frac{V_\text{c}}{-V_{,\phi_\text{c}}} \Big) \frac{1}{|H_\text{c}| \sqrt{2 X_\text{c}}} \Big| \frac{\Delta X}{2 X_\text{c}} \Big| \ll I(t_\text{c}) .
\end{align}
Therefore it simply contributes a constant factor to the outer integral in (\ref{eq:psis-const+int}), which leads to
\begin{align} \label{eq:psis-int}
\psi_\text{s}^\text{int} &\approx C_2 (k) \int_0^t dt' \dot{H} \, \frac{1}{3 a^3_\text{ek-end}} \Big( \frac{V_\text{c}}{- V_{,\phi_\text{c}}} \Big) \frac{1}{|H_\text{c}| \sqrt{2 X_\text{c}}} \nonumber \\
&\approx \frac{C_2 (k)}{3 a^3_\text{ek-end}} \Big( \frac{V_\text{c}}{- V_{,\phi_\text{c}}} \Big) \frac{1}{\sqrt{2 X_\text{c}}} \int_{H_\text{c}}^H \frac{dH}{|H_\text{c}|} \, .
\end{align}
Clearly when $H \rightarrow 0$ near the bounce, $\psi_\text{s}^\text{int}$ approaches the same value as $R^\text{int}$ given by Eq.~(\ref{eq:Rw-1}), which is exponentially large and nonlinear for a wide range of $k$ modes. From the above derivation we also see that $\psi_\text{s}$ becomes exponentially large only when $|H| \ll |H_\text{c}|$ near the bounce. This corresponds to $|t - t_\text{0}| \ll 1 / |H_\text{c}|$ in the bouncing solution (\ref{eq:bounceH}), which is within the last e-fold before the bounce. We shall see below that during this time the anisotropy grows large and nonlinear, too.

\section{Anisotropy and nonsingular bounce} \label{sec:anisotropy}

Anisotropy is described by the shear of the constant time hypersurface. In Appendix~\ref{sec:hypersurface} we derive the generalized local Friedmann equations, and show that the anisotropy term exactly corresponds to the squared magnitude of the shear. Here we illustrate a simpler case with a flat, homogeneous, but anisotropic metric,
\begin{equation}
ds^2 = - \mathcal{N}^2 d\tau^2 + a^2 \, \tilde{\gamma}_{ij} ( dx^i + \beta^i d\tau ) ( dx^j + \beta^j d\tau ) ,
\end{equation}
where the lapse $\mathcal{N}$, the shift $a^2 \beta^i$, and the spatial metric $a^2 \tilde{\gamma}_{ij}$ depend only on $\tau$. We choose the scale factor $a(\tau)$ such that $\det \tilde{\gamma}_{ij} = 1$. The extrinsic curvature of the constant time hypersurface with normal $n_\mu = (-\mathcal{N}, \vec{0}\,)$ is given by
\begin{equation}
K_{ij} = n_{j;i} = \tfrac{1}{3} \theta \gamma_{ij} + \sigma_{ij} \ ,
\end{equation}
where it is decomposed into the expansion $\theta$ and the shear $\sigma_{ij}$ by using the projection tensor $\gamma_{\mu\nu} = g_{\mu\nu} + n_\mu n_\nu$,
\begin{align}
\theta &\equiv n^i_{\; ;i} = \frac{3 \mathcal{H}}{\mathcal{N}} \ , \\
\sigma_{ij} &\equiv n_{i;j} - \tfrac{1}{3} \gamma_{ij} n^k_{\; ;k} = \frac{a^2}{2 \mathcal{N}} \tilde{\gamma}_{ij}^{\,\prime} \ ,
\end{align}
and $\mathcal{H} \equiv \frac{a'}{a}$ is the conformal Hubble parameter. The squared magnitude of the shear is given by
\begin{equation}
\sigma^2 \equiv \tfrac{1}{2} \sigma^{ij} \sigma_{ij} = \frac{1}{8 \mathcal{N}^2} \tilde{\gamma}^{ij} \tilde{\gamma}_{jk}^{\,\prime} \tilde{\gamma}^{k\ell} \tilde{\gamma}_{\ell i}^{\,\prime} \ .
\end{equation}
For a perfect fluid at rest, {\it e.g.} a homogeneous scalar field, the Einstein equations can be written as
\begin{align}
\rho &= - G^0_{\; 0} = \frac{1}{3} \theta^2 - \sigma^2 \ , \\[4pt]
3p &= G^i_{\; i} = - \frac{2}{\mathcal{N}} \theta^\prime - \theta^2 - 3 \sigma^2 \ , \\[4pt]
0 &= G^i_{\; j} - \tfrac{1}{3} \delta^i_{\, j} G^k_{\; k} = \frac{1}{\mathcal{N}} \sigma^i_{\; j} {}^\prime + \theta \sigma^i_{\; j} \ .
\end{align}
In physical time $dt = \mathcal{N} d\tau$, the Hubble parameter is $H = \mathcal{H} / \mathcal{N} = \theta / 3$, and the first two equations reduce to the Friedmann equations,
\begin{align}
H^2 &= \tfrac{1}{3} ( \rho + \sigma^2 ) \; , \label{eq:Friedmann1-homo} \\[4pt]
\dot{H} &= - \tfrac{1}{2} ( \rho + p ) - \sigma^2 \; . \label{eq:Friedmann2-homo}
\end{align}
The third Einstein equation becomes
\begin{equation}
\dot{\sigma}^i_{\; j} + 3 H \sigma^i_{\; j} = 0,
\end{equation}
which implies that the shear scales as
\begin{equation} \label{eq:sigmaproptoa3}
\sigma^i_{\; j} \propto \frac{1}{a^3} \; , \quad \sigma^2 \propto \frac{1}{a^6} \; .
\end{equation}
Thus the anisotropy term in the Friedmann equation (\ref{eq:Friedmann1}) is precisely the squared shear of the constant time hypersurface. More general cases with nonflat and inhomogeneous metrics are dicussed in Appendix~\ref{sec:hypersurface}.

For our analysis of the bouncing dynamics, we again consider the perturbatively inhomogeneous and anisotropic metric (\ref{eq:pFRW}),
\begin{align*}
ds^2 = &\ a^2 \Big[ - ( 1 + 2 A) d\tau^2 + 2 ( B_{,i} + S_i ) d\tau dx^i \\
& + \big( ( 1 - 2 \psi ) \delta_{ij} + 2 E_{,ij} + 2 F_{(i,j)} + 2 h_{ij} \big) dx^i dx^j \Big] . \nonumber
\end{align*}
Small anisotropies can be analyzed by studying the shear perturbation of the constant time hypersurfaces. The normal to these hypersurfaces is $n_\mu = (-a(1+A), \vec{0}\,)$, and the shear is given by (Appendix (\ref{eq:sigma_pert}))
\begin{align}
\sigma_{ij} = & \; a \big( (E^\prime_{,ij} - B_{,ij}) - \tfrac{1}{3} \delta_{ij} \nabla^2 (E^\prime - B) \big) \nonumber \\
& + a \big( F_{(i,j)}^\prime - S_{(i,j)} \big) .
\end{align}
It involves both the scalar and the vector perturbations, which evolve independently at linear order. The scalar part of the shear is coupled to the curvature perturbation through Einstein equations, while the vector part scales in a simple manner with the scale factor $a$. Therefore, we expect the anisotropy to grow significantly whenever the curvature perturbation becomes large, or when the scale factor shrinks exponentially as in the bouncing phase described in Section~\ref{sec:bounce}. In that case, the rise of a large anisotropy would substantially alter the dynamics of the contracting Universe and potentially demolish the nonsingular bounce.

Let us first look at the vector perturbations that follows very simple behavior. The vector part of the shear perturbation is given by
\begin{equation}
\sigma^\text{V}_{ij} = a \big( F_{(i,j)}^\prime - S_{(i,j)} \big) \equiv a \, \sigma^\text{V}_{(i,j)} \ ,
\end{equation}
where
\begin{equation}
\sigma^\text{V}_i \equiv F_i^\prime - S_i \ ,
\end{equation}
and $\sigma^\text{V} {}^i_{\; ,i} = 0$. The variable $\sigma^\text{V}_i$ is gauge-invariant by itself, and evolves according to the equation (Appendix (\ref{eq:sigmaVecij}))
\begin{equation}
\sigma^\text{V}_i {}^\prime + 2 \mathcal{H} \, \sigma^\text{V}_i = 0 ,
\end{equation}
which implies
\begin{equation} \label{eq:sigmaVi}
\sigma^\text{V}_i \propto \frac{1}{a^2} \, ,
\end{equation}
and hence
\begin{equation} \label{eq:sigmaVij}
\sigma^\text{V} {}^i_{\;j} = \frac{1}{a^2} \, \sigma^\text{V}_{ij} = \frac{1}{a} \, \sigma^\text{V}_{(i,j)} \propto \frac{1}{a^3} \, ,
\end{equation}
in agreement with (\ref{eq:sigmaproptoa3}). Unlike in an expanding universe, the vector perturbation $\sigma^\text{V}_i$, if nonzero, grows in the contracting universe as the scale factor $a$ decreases. Therefore, in general, the vector perturbation should not be overlooked in analyzing bouncing cosmologies \cite{Battefeld:2004cd}, since it naturally creates anisotropy and may disrupt the bounce. As we argued in Section~\ref{sec:bounce}, the exponential growth of anisotropy in the bouncing phase may cause the breakdown of perturbation theory near the bounce. However, since the vector perturbation is not sourced by the scalar field, it will remain negligible if it is extremely small initially. In this sense, the problem with the vector part of the anisotropy can be resolved by fine-tuning initial conditions.

Suppose that the universe is homogeneous but there exists some initial anisotropy $\sigma^2_\text{ek-beg}$ at the beginning of the ekpyrotic phase. We would like to determine if, after exponential amplification in the bouncing phase, it may dominate the energy density and prevent the nonsingular bounce. For this purpose, we shall include the anisotropy term in the Friedmann equations and reanalyze the bouncing behavior of the Universe. Under the approximations (\ref{eq:approxT}) and (\ref{eq:approxV}), the Friedmann equations become
\begin{align}
3 H^2 &= \rho_\phi + \sigma^2 \approx T + V_\text{c} + \sigma^2 , \label{eq:Friedmann1new} \\
\dot{H} &= - X P_{,X} - \sigma^2 \approx \frac{-T}{2} - \sigma^2 . \label{eq:Friedmann2new}
\end{align}
In order to reach a nonsingular bounce in the presence of a nonzero $\sigma^2$, $\rho_\phi$ must decrease to a negative value to cancel the $\sigma^2$ in (\ref{eq:Friedmann1new}). Meanwhile, $\dot{H}$ must stay positive before $H$ hits zero; hence, the kinetic energy $T$ has to be sufficiently negative so that its magnitude is larger than $2 \sigma^2$ in (\ref{eq:Friedmann2new}). From Eq.~(\ref{eq:bouncea}) we know that in the bouncing phase $T$ scales as $1/a^3$, slower than $\sigma^2$ which grows as $1/a^6$. Therefore if $\sigma^2$ starts with too large an initial value, it may overtake the scalar field energy and prevent the bounce.

Here we derive a general condition for the nonsingular bounce to happen. Our only assumption is that $X$, and hence $T$, is monotonically decreasing with time in the bouncing phase, which is guaranteed by the steep potential that slows down the field $\phi$. Let the anisotropy be $\sigma^2_{1}$ at some time $t_1$ when it is still much smaller than the scalar field energy $\rho_1$. Introduce the variable
\begin{equation}
q \equiv \Big( \frac{a_1}{a} \Big)^3 ,
\end{equation}
which starts from $q=1$ and increases as $a$ decreases. Since $\sigma^2 \propto a^{-6} \propto q^2$, we have
\begin{equation}
\frac{d \sigma^2}{dq} = \sigma^2_1 \cdot 2 q .
\end{equation}
For the scalar field, from Eq.~(\ref{eq:EOM}), we have
\begin{equation}
\dot{\rho_\phi} + 3 H T = 0 ,
\end{equation}
and hence
\begin{equation}
\frac{d \rho_\phi}{dq} = \frac{T}{q} \leq \frac{T_1}{q} ,
\end{equation}
since $T$ is monotone decreasing. Therefore, Eq.~(\ref{eq:Friedmann1new}) satisfies
\begin{equation}
\frac{d (3 H^2)}{dq} \leq - \frac{|T_1|}{q} + 2 \sigma^2_1 \, q ,
\end{equation}
which integrates to give
\begin{equation}
3 H^2 \leq 3 H^2_1 - |T_1| \, \log q + \sigma^2_1 (q^2 - 1) .
\end{equation}
To have a bounce at finite $a$, or finite $q$, a sufficient condition is that the RHS has a root in $q > 1$, or equivalently, the minimum of the RHS is less than $0$. The minimum is at $q^2 = |T_1| / 2 \sigma^2_1$, where the RHS equals
\begin{equation*}
\rho_1 + \frac{|T_1|}{2} - \frac{|T_1|}{2} \log \frac{|T_1|}{2 \sigma^2_1} \, .
\end{equation*}
Requiring this to be less than $0$, we find the condition
\begin{equation} \label{eq:sigma1cond}
\sigma^2_1 \leq \frac{|T_1|}{2} \exp \{ - \tfrac{2 \rho_1}{|T_1|} - 1 \} .
\end{equation}

This condition applies for any point $t_1$ in the bouncing phase, in particular the last stage where Eq.~(\ref{eq:EOMT}) and hence $T \propto 1/a^3$ does not hold. Nonetheless, since it is sufficient at any point $t_1$, we may find the mildest condition by a proper choice of $t_1$. Let $t_1$ be within the middle stage where we can use $T \propto 1/a^3$. Extrapolating back to the beginning of the bouncing phase, we need to have
\begin{align}
\sigma^2_\text{bp-beg} &\leq \frac{a_1^6}{a_\text{bp-beg}^6} \frac{|T_1|}{2} \exp \{ - \tfrac{2 (V_\text{c} - |T_1|)}{|T_1|} - 1 \} \nonumber \\
&= \frac{|T_\text{bp-beg}|^2}{2 |T_1|} \exp \{ 1 - \tfrac{2 V_\text{c}}{|T_1|} \} .
\end{align}
The mildest condition is established at $t_1 \rightarrow t_\text{bp-end}$ where $|T_1| \rightarrow |T_\text{bp-end}| \approx V_\text{c}$, which imposes
\begin{equation}
\sigma^2_\text{bp-beg} \leq \frac{|T_\text{bp-beg}|^2}{2 V_\text{c}} e^{-1} \approx \Big( \frac{- V_{,\phi_\text{c}}}{V_\text{c}} \Big)^{\! 2} \, \frac{X_\text{c}}{3 e} \sim X_\text{c} \, .
\end{equation}
Therefore, since $\sigma^2$ is nearly constant through the ekpyrotic phase, the condition on the initial value of $\sigma^2_\text{ek-beg}$ is simply
\begin{equation} \label{eq:initAnis}
\sigma^2_\text{ek-beg} \lesssim X_\text{c} \; ,
\end{equation}
which has to be much less than ${\rho_\phi}_\text{ek-beg} \approx 3 p X_\text{ek-beg}$. In other words, to prevent the anisotropy from becoming a problem during the bouncing phase, the initial conditions must be fine-tuned so that the ekpyrotic phase is highly isotropic to begin with.

Condition (\ref{eq:initAnis}) applies for not only the vector perturbation but any initial anisotropy that scales as $1/a^6$, including any initial scalar contributions. However, for scalar perturbations, large anisotropy can arise even if there is no initial anisotropies classically, because the scalar perturbation is continuously sourced by the quantum fluctuations of the scalar field. We demonstrate that the scalar shear perturbation is induced during the ekpyrotic phase and grows exponentially large near the bounce.

The scalar part of the shear perturbation is given by
\begin{align}
\sigma^\text{S}_{ij} &= a \big( (E^\prime_{,ij} - B_{,ij}) - \tfrac{1}{3} \delta_{ij} \nabla^2 (E^\prime - B) \big) \nonumber \\
&\equiv a \big( \sigma^\text{S}_{,ij} - \tfrac{1}{3} \delta_{ij} \nabla^2 \sigma^\text{S} \big) ,
\end{align}
where
\begin{equation}
\sigma^\text{S} \equiv E^\prime - B \; .
\end{equation}
Under the coordinate transformation $x^\mu \rightarrow x^\mu + \xi^\mu$, $\sigma^\text{S}$ transforms as (Appendix (\ref{eq:sigma_transf}))
\begin{equation}
\sigma^\text{S} \rightarrow \sigma^\text{S} - \xi^0 \; ,
\end{equation}
and thus is not gauge-invariant. In the comoving gauge, $\sigma^\text{S}$ is described by the gauge-invariant quantity
\begin{equation}
\sigma_\text{c} \equiv \sigma^\text{S} - \frac{\delta \phi}{\phi '} \; .
\end{equation}
The evolution of $\sigma_\text{c}$ is coupled to the comoving curvature perturbation $\mathcal{R}$ through the equations (Appendix (\ref{eq:G00com}$\sim$\ref{eq:Gijcom}))
\begin{align}
& \sigma_\text{c}^\prime + 2 \mathcal{H} \sigma_\text{c} + \frac{\mathcal{R}^\prime}{\mathcal{H}} + \mathcal{R} = 0 , \\
& \mathcal{R}^\prime + \frac{c_s^2 \mathcal{H}}{\mathcal{H}^\prime - \mathcal{H}^2} \nabla^2 ( \mathcal{R} + \mathcal{H} \sigma_\text{c} ) = 0 .
\end{align}
Thereby the scalar shear and curvature perturbations source each other to grow from quantum fluctuations to classical perturbations when the modes exit the horizon. From the second equation above we can solve for $\sigma_\text{c}$ in terms of $\mathcal{R}$,
\begin{equation}
\sigma_\text{c} = \frac{\mathcal{H}^\prime - \mathcal{H}^2}{c_s^2 k^2 \mathcal{H}^2} \mathcal{R}^\prime - \frac{\mathcal{R}}{\mathcal{H}} \; .
\end{equation}
Since we know from Section~\ref{sec:curvature} that the comoving curvature perturbation $\mathcal{R}$ undergoes exponential amplification right before the bouncing phase when $w$ crosses $-1$, we expect the comoving shear perturbation $\sigma_\text{c}$ to follow the same growth. Indeed, using the leading order solution (\ref{eq:R0}) plus the expansion (\ref{eq:Rexpansion}) of $\mathcal{R}$, we find, up to order $\mathcal{O} (k^{1/2})$,
\begin{align} \label{eq:sigmacint}
\sigma_\text{c} = & - \frac{C_2 (k)}{2 k^2 a^2} - \frac{C_1 (k)}{a^2} \int_0^t \! dt' a \\
& + \frac{C_2 (k)}{a^2} \bigg( \frac{a}{H} \! \int_0^t \!\! dt' \frac{c_s^2 H^2}{2 a^3 \dot{H}} + \int_0^t \!\! dt' \frac{a \dot{H}}{H^2} \int_0^{t'} \!\!\! dt'' \frac{c_s^2 H^2}{2 a^3 \dot{H}} \bigg) . \nonumber
\end{align}
Along the same lines as for Eqs.~(\ref{eq:Rw-1}) and (\ref{eq:psis}), we see that the third term increases exponentially as $\dot{H} \rightarrow 0$, while the fourth and all higher order terms are well behaved and finite. Compared to the leading term, $\sigma_\text{c}$ becomes dominated by the third term which surpasses the leading term by a factor
\begin{equation} \label{eq:sigmac_w-1}
\frac{\sigma_\text{c} |_{w \rightarrow -1}}{\sigma_\text{c} |_{\text{ek-end}}} \approx \frac{2 k^2}{3 (aH)_\text{ek-end}^2} \Big( \frac{V_\text{c}}{- V_{,\phi_\text{c}}} \Big) \sqrt{\frac{p X_\text{ek-end}}{2 X_\text{c}}} \sim e^{N - 2 N_k} .
\end{equation}
Note that this is the same exponential factor as in (\ref{eq:Rw-1toEntro}), by which the comoving curvature perturbation $\mathcal{R}$ grows when $w \rightarrow -1$ for the range of modes with $N_k < N/2$.

However, this huge growth of the comoving shear perturbation does not imply the breakdown of perturbation theory at this point. As we have seen in Section~\ref{sec:curvature}, in the synchronous gauge the curvature perturbation $\psi_\text{s}$ remains small throughout the contracting phase until very near the bounce. Accordingly, the shear perturbation would also remain small in this gauge. The synchronous shear perturbation $\sigma_\text{s}$ can be related to the comoving shear perturbation $\sigma_\text{c}$ through the gauge transformation (Appendix (\ref{eq:sigma_c-to-sigma_s}))
\begin{equation}
\sigma_\text{s} = \sigma_\text{c} + \frac{1}{a} \int_0^t \frac{\dot{\mathcal{R}}}{H} dt' \; .
\end{equation}
After the mode exits the horizon, we use Eqs.~(\ref{eq:sigmacint}) and (\ref{eq:Rconst+Rint}) to find, up to order $\mathcal{O}(k^{1/2})$,
\begin{align}
\sigma_\text{s} = & - \frac{C_2 (k)}{2 k^2 a^2} - \frac{C_1 (k)}{a^2} \int_0^t \! dt' a + \frac{C_2 (k)}{a^2} \bigg( \frac{a}{H} \int_0^t \!\! dt' \frac{c_s^2 H^2}{2 a^3 \dot{H}} \nonumber \\
& + \int_0^t \!\! dt' \frac{a \dot{H}}{H^2} \int_0^{t'} \!\!\! dt'' \frac{c_s^2 H^2}{2 a^3 \dot{H}} \bigg) - \frac{C_2 (k)}{a} \int_0^t \!\! dt' \frac{c_s^2 H}{2 a^3 \dot{H}} \nonumber \\
= & - \frac{C_2 (k)}{2 k^2 a^2} - \frac{C_1 (k)}{a^2} \int_0^t \! dt' a + \frac{C_2 (k)}{a^2} \times \nonumber \\
& \bigg( \int_0^t \!\! dt' \frac{a \dot{H}}{H^2} \int_0^{t'} \!\!\! dt'' \frac{c_s^2 H^2}{2 a^3 \dot{H}} - a \int_0^t \!\! dt' \frac{\dot{H}}{H^2} \int_0^{t'} \!\!\! dt'' \frac{c_s^2 H^2}{2 a^3 \dot{H}} \bigg) \nonumber \\
= & - \frac{C_2 (k)}{2 k^2 a^2} - \frac{C_1 (k)}{a^2} \int_0^t \! dt' a \nonumber \\
& - \frac{C_2 (k)}{a^2} \int_0^t \!\! dt' a H \int_0^{t'} \!\!\! dt'' \frac{\dot{H}}{H^2} \int_0^{t''} \!\!\!\! dt''' \frac{c_s^2 H^2}{2 a^3 \dot{H}} \, .
\end{align}
From the first to the second line it is clear that the growing term in $\sigma_\text{c}$ is absorbed by the gauge transformation term, leaving behind an integral that is well behaved as $\dot{H} \rightarrow 0$, similar to Eq.~(\ref{eq:psis}).

Moreover, using similar calculations that lead to (\ref{eq:psis-int}), $\sigma_\text{s}$ in the bouncing phase can be evaluated as
\begin{align} \label{eq:sigmas-bp}
\sigma_\text{s} \approx & - \frac{C_2 (k)}{2 k^2 a^2} - \frac{C_1 (k)}{a^2} \int_0^t \! dt' a \nonumber \\
& - \frac{C_2 (k)}{a^2} \int_0^t \! dt' a H \cdot \frac{1}{H} \frac{1}{3 a^3_\text{ek-end}} \Big( \frac{V_\text{c}}{- V_{,\phi_\text{c}}} \Big) \frac{1}{\sqrt{2 X_\text{c}}} \nonumber \\
\approx & - \frac{C_2 (k)}{2 k^2 a^2} - \bigg[ C_1 (k) + \frac{C_2 (k)}{3 a^3_\text{ek-end}} \frac{( - V_\text{c} / V_{,\phi_\text{c}} )}{\sqrt{2 X_\text{c}}} \bigg] \frac{1}{a^2} \! \int_0^t \!\! dt' a \nonumber \\
\approx & - \frac{C_2 (k)}{2 k^2 a^2} - \frac{C_1 (k)}{a^2} \frac{a_\text{bp-beg}}{|H_\text{c}|} - \frac{1}{a^2} \frac{C_2 (k)}{3 a^2_\text{ek-end}} \frac{( - V_\text{c} / V_{,\phi_\text{c}} )}{|H_\text{c}| \sqrt{2 X_\text{c}}} \, .
\end{align}
The integral in the second line is found by using (\ref{eq:bouncea}), which rapidly converges to the constant value $\approx a_\text{bp-beg} / |H_\text{c}|$. From the last line we see that all three terms scale equally as $1/a^2$ near the bounce, which coincides with the vector perturbation (\ref{eq:sigmaVi}) and likewise leads to
\begin{equation} \label{eq:sigmasij}
\sigma_\text{s} {}^i_{\;j} = \frac{1}{a} \big( \sigma_\text{s} {}_{,ij} - \tfrac{1}{3} \delta_{ij} \nabla^2 \sigma_\text{s} \big) \propto \frac{1}{a^3}
\end{equation}
that resembles Eq.~(\ref{eq:sigmaproptoa3}) in the homogeneous case. We also find that the third term in (\ref{eq:sigmas-bp}) dominates over the first (and second) term by a ratio
\begin{align}
\frac{2 k^2}{3 a^2_\text{ek-end}} \frac{( - V_\text{c} / V_{,\phi_\text{c}} )}{|H_\text{c} |\sqrt{2 X_\text{c}}} \sim \frac{k^2}{(aH)_\text{ek-end}^2} \sqrt{\frac{X_\text{ek-end}}{X_\text{c}}} \sim e^{N - 2 N_k}
\end{align}
for modes with $N_k < N/2$, similar to (\ref{eq:sigmac_w-1}) but only at a much later time in the bouncing phase.

We note that the scalar shear perturbation is generated by quantum fluctuations and cannot be fine-tuned away. To see whether the resulting anisotropy grows nonlinear, we estimate the size of the anisotropy by computing the self-correlation function of the scalar shear perturbation,
\begin{align} \label{eq:<sigma2>}
\langle (\sigma^\text{S})^2 \rangle &= \langle \tfrac{1}{2} \, \sigma^\text{S} {}^{ij} (\vec{x}) \, \sigma^\text{S}_{\; ij} (\vec{x}) \rangle \nonumber \\
&= \int \frac{d^3 k}{(2\pi)^3} \, \frac{1}{2 a^2} \Big| \big( - k_i k_j + \tfrac{1}{3} \delta_{ij} k^2 \big) \sigma^\text{S}_{\; k} \Big|^2 \nonumber \\
&= \int \frac{1}{6 \pi^2 a^2} \, |\sigma^\text{S}_{\; k}|^2 \, k^6 dk \; .
\end{align}
The integration is carried over the modes that exit the horizon in the second half of the ekpyrotic phase, {\it i.e.} those that have $0 < N_k < N/2$ and are dominated by the third term in (\ref{eq:sigmas-bp}). Modes with $N_k > N/2$ are negligible since the power spectrum of $\sigma^\text{S}$ is deeply blue, whereas the modes with $N_k < 0$ remain inside the horizon until very close to the bounce, contributing only vacuum fluctuations that can be eliminated by renormalization. Therefore, as we approach the putative bounce, a lower estimate of the anisotropy is
\begin{align} \label{eq:sigma2-bpend}
\langle (\sigma^\text{S})^2 \rangle &\gtrsim \int\limits_{e^{- ^N \!\! / \! _2} (aH)_\text{ek-end}}^{(aH)_\text{ek-end}} \hspace{-5pt} \frac{1}{6 \pi^2 a^6} \Big| \frac{C_2 (k)}{3 a_\text{ek-end}^2} \frac{( - V_\text{c} / V_{,\phi_\text{c}} )}{|H_\text{c} |\sqrt{2 X_\text{c}}} \Big|^2 \, k^6 dk \nonumber \\
&\approx \frac{1}{108 \pi^2 a^6} \frac{( - V_\text{c} / V_{,\phi_\text{c}} )^2}{a_\text{ek-end}^4 H_\text{c}^2 X_\text{c}} \negthickspace \int\limits^{(aH)_\text{ek-end}} \hspace{-15pt} |C_2 (k)|^2 k^6 dk \nonumber \\
&\approx \frac{1}{108 \pi^2 a^6} \frac{( - V_\text{c} / V_{,\phi_\text{c}} )^2}{a_\text{ek-end}^4 H_\text{ek-end}^2 X_\text{c}} \; \frac{a_\text{ek-end}^2}{8p} (aH)_\text{ek-end}^8 \nonumber \\
&\sim \frac{a_\text{ek-end}^6}{a^6} \Big( \frac{V_\text{c}}{- V_{,\phi_\text{c}}} \Big)^{\! 2} \, \frac{X_\text{ek-end}}{X_\text{c}} H_\text{ek-end}^4 \; ,
\end{align}
where we used the expression (\ref{eq:C2}) for $C_2(k)$. At $t \rightarrow t_\text{bp-end}$, we have
\begin{equation}
\langle (\sigma^\text{S})^2 \rangle _\text{bp-end} \gtrsim \frac{a_\text{bp-beg}^6}{a_\text{bp-end}^6} \, e^{2N} \, H_\text{ek-end}^4 \sim V_\text{c}^2 \, e^{4N} .
\end{equation}
In order for the anisotropy not to become a problem, condition (\ref{eq:sigma1cond}) requires that
\begin{equation} \label{eq:sigma2-bpend-cond}
\sigma_\text{bp-end}^2 \leq \frac{V_\text{c}}{2 e} \, .
\end{equation}
This can be satisfied only if $V_\text{c}^2 \, e^{4N} \lesssim V_\text{c}$ , or
\begin{equation} \label{eq:cond_Vc}
V_\text{c} \lesssim e^{-4N} M_\text{Pl}^4
\end{equation}
in reduced Planck units. However, in order for the scale-invariant modes to match the observed amplitude $\frac{\delta \rho_k}{\rho} \sim 10^{-5}$, the potential energy at the end of the ekpyrotic phase needs to satisfy $\sqrt{2 V_\text{ek-end} / p} \sim 10^{-3} M_\text{Pl}^2$ \cite{Lehners:2007ac}, which implies
\begin{equation} \label{eq:Vc}
V_\text{c} \approx 3 p V_\text{ek-end} \sim p^2 10^{-6} M_\text{Pl}^4 \; .
\end{equation}
For a typical value of $p \sim 10^{-2}$ \cite{Khoury:2003rt}, the condition (\ref{eq:cond_Vc}) is incredibly far from being satisfied.

Since the scalar shear perturbation $\sigma^\text{S}$ is not gauge-invariant, we must explain our gauge choice for computing (\ref{eq:sigma2-bpend}). Recall that condition (\ref{eq:sigma2-bpend-cond}) is obtained for the homogeneous case by using the Friedmann equations of the Hubble parameter $H$. In the inhomogeneous case, the Friedmann equations should be generalized to equations of the local expansion $\theta$ of the constant time hypersurface (see (\ref{eq:localFriedmann1}, \ref{eq:localFriedmann2}) in Appendix~\ref{sec:hypersurface}). In general, the expansion $\theta$ is not uniform on the hypersurface, and its gradient along the hypersurface is determined by a supplementary equation (\ref{eq:localFriedmann3}). In order to apply condition (\ref{eq:sigma2-bpend-cond}), it is preferable to adopt the constant mean curvature gauge described in Appendix~\ref{sec:hypersurface}. In this gauge Eq.~(\ref{eq:localFriedmann3}) vanishes identically, so that $\theta$ is indeed uniform and equal to $3H$ on the constant time hypersurface. This gauge coincides with the synchronous gauge in the homogeneous case. For linear perturbations, this gauge reduces to the uniform Hubble gauge defined in Appendix~\ref{sec:perturbation}. The scalar shear perturbation in the uniform Hubble gauge can be computed through (\ref{eq:sigmac-to-sigmaH}) and is found to be equal to the synchronous shear perturbation $\sigma_\text{s}$ to leading order, which is what we used to compute Eq.~(\ref{eq:sigma2-bpend}) above.

It is worth contrasting the uniform Hubble gauge with the longitudinal gauge in which the scalar shear perturbation $\sigma^\text{S}$ is set to zero identically. The vanishing shear perturbation in the longitudinal gauge does not mean that the bouncing phase is free from problems. According to Eqs.~(\ref{eq:deltaH}) and (\ref{eq:G0i}), the variation of the local Hubble parameter over the constant time hypersurface is given by
\begin{equation}
\delta \mathcal{H} = - ( \mathcal{H}^\prime \! - \! \mathcal{H}^2 ) \delta u + \tfrac{1}{3} \nabla^2 \sigma^\text{S} \; .
\end{equation}
In the longitudinal gauge, though $\sigma^\text{S} = 0$, the velocity perturbation $\overline{\delta u} = \sigma_\text{c}$ (Appendix~(\ref{eq:deltauNewt})) grows exponentially due to Eq.~(\ref{eq:sigmac_w-1}). Therefore the local Hubble parameter is extremely inhomogeneous over the hypersurface, which means the nonsingular bounce is ill defined in this gauge even when the background Hubble parameter $H$ vanishes. In the uniform Hubble gauge, however, $\delta \mathcal{H} = 0$ on every constant time hypersurface, hence the bounce would be reached simultaneously at the hypersurface on which $H = \theta/3 = 0$. Therefore the nonsingular bounce is well defined in this gauge, and the huge growth of anisotropy is indeed physical.

In a consistent perturbative analysis, the $\sigma^2$ term in the Friedmann equations should be negligible to linear order. The exponentially large $\langle (\sigma^\text{S})^2 \rangle$ that we found in (\ref{eq:sigma2-bpend}) implies that the perturbation theory is no longer valid. Therefore we expect the anisotropy generated by quantum fluctuations to become nonlinear and dominate over the scalar field energy before the bounce. Once the anisotropy takes over, the Universe will be driven to a BKL like contraction phase that ends in an extremely inhomogeneous and anisotropic singular crunch \cite{Misner:1969hg, Belinsky:1970ew}.

\section{Conclusion} \label{sec:conclusion}

We have examined a cosmological model of a contracting universe that smoothly connects an ekpyrotic phase with a nonsingular bounce. We have shown how the nonsingular bouncing phase spoils the flatness, homogeneity, and isotropy condition, as well as the scale-invariant perturbations generated during the ekpyrotic phase. We have identified four different effects that create problems for nonsingular bouncing models: \\
$\bullet$ \textbf{gravitational instability:} This problem is caused by $c_s^2$ becoming negative in the bouncing phase. One solution is to have $c_s^2$ remain exponentially small during the bouncing phase, as in (\ref{eq:cs2cond}). Alternatively, it has been suggested \cite{Creminelli:2006xe, Buchbinder:2007ad} to restrict the bouncing phase to one Hubble time; this requires tuning $(- V_{,\phi_\text{c}} / V_\text{c})$ to be exponentially large, which is not viable, as shown below. \\
$\bullet$ \textbf{regrowth of initial anisotropy:} The initial anisotropy suppressed during the ekpyrotic phase is exponentially amplified to even greater magnitudes during the bouncing phase according to Eq.~(\ref{eq:sigmabpendtobeg}). This creates a problem for the nonsingular bounce unless the initial anisotropy present before the ekpyrotic phase is fine-tuned to an extremely small value given by (\ref{eq:initAnis}), {\it e.g.} by having a dark energy dominated expanding phase preceding the ekpyrotic phase, as in \cite{Steinhardt:2001st, Steinhardt:2002ih}. \\
$\bullet$ \textbf{blue spectrum of curvature perturbations:} As first discovered in \cite{Xue:2010ux}, a sub-subdominant component of adiabatic curvature perturbation generated from quantum fluctuations in the ekpyrotic phase grows to dominate over the scale-invariant component as the bouncing phase begins, resulting in a blue power spectrum that is inconsistent with observations. This problem occurs when the equation of state $w$ passes through $-1$, and it persists even if the gravitational instability is suppressed. \\
$\bullet$ \textbf{growth of quantum induced anisotropy:} The scalar shear perturbation generated from quantum fluctuations in the ekpyrotic phase grows exponentially during the bouncing phase and contributes a large anisotropy that dominates the energy density and prevents the nonsingular bounce. This problem is due to both the equation of state $w$ passing through $-1$ and the scale factor $a$ decreasing exponentially in the bouncing phase. It persists even if the first two problems are avoided.

According to Eqs.~(\ref{eq:Rw-1toEnd}), (\ref{eq:abp-end}), and (\ref{eq:sigma2-bpend}), it would appear that the last two problems could be resolved if the factor $( - V_{,\phi_\text{c}} / V_\text{c} )$ is made to be at least of order $e^N$; the first two problems would not occur either if this condition were achieved, which amounts to keeping the duration of the bouncing phase within one Hubble time. However, this condition requires $| V_{,\phi_\text{c}} | \gtrsim V_\text{c} \, e^{N}$, and by Eq.~(\ref{eq:Vc}) the value of $|V_{,\phi_\text{c}}|$ would exponentially exceed the Planck scale. It is then unavoidable to consider quantum gravity effects in this approach, which defeats the purpose of the nonsingular bounce to avoid those effects.

It is worth noting that none of the above four problems appear in the singular bounce, as in the cyclic model \cite{Steinhardt:2001st, Steinhardt:2002ih}. The first two problems disappear because in the singular bounce $c_s^2$ is always positive and the horizon is forever shrinking. Moreover, in the singular case, $w$ remains $> 1$ throughout the contraction phase and the energy density is increasing all the way to the bounce, so the curvature and anisotropy never grow to dominate. But in the nonsingular approach considered here, $w$ must fall below $-1$ and $X$ must fall below $X_\text{c}$, and it is during this phase that the curvature and shear perturbations grow to make the universe inhomogeneous and anisotropic.

In the introduction to this paper, we noted a decoupling argument suggesting that the bouncing phase should not affect the large scale density perturbations or the isotropy of the Universe, as long as the bounce occurs at much higher energy densities than during the ekpyrotic phase when the perturbations are first generated. Our conclusion does not contradict this intuition, but instead supports it: the failure of the nonsingular bounce considered here is precisely due to the fact that the bounce is mediated by ghost condensation, a low energy effect rather than a high energy one.

Indeed, from Eqs.~(\ref{eq:Rw-1toEnd}) and (\ref{eq:sigmabpendtobeg}) it is clear that both growths of curvature and anisotropy are proportional to the same factor $X_\text{ek-end} / X_\text{c}$, the ratio between the high energy scale near the end of the ekpyrotic phase and the low energy scale associated with the ghost condensation. In order to consistently merge the ekpyrotic phase with the bouncing phase through ghost condensation, these energy scales have to satisfy the hierarchy
\begin{equation} \label{eq:hierarchy}
X_\text{c} \ll X_\text{ek-beg} \ll X_\text{ek-end} \; .
\end{equation}
Therefore the exponentially large factor is inherent in such ekpyrotic nonsingular bouncing models. Our calculation illustrates how this same factor naturally shows up and leads to the four problems listed above.

Therefore, we conjecture that any form of nonsingular bouncing models would suffer from these problems unless the hierarchy (\ref{eq:hierarchy}) can be relaxed. This may happen either by reducing the energy ratio between the beginning and the end of the ekpyrotic phase, or by raising the energy scale of the bounce. The first possibility is to seek a nonlinear realization of the ekpyrotic phase in which $X$ varies within a limited range. The other possibility is to induce a bounce with physics of a much higher energy scale. We are currently attempting to construct examples of both types.

\begin{acknowledgments}

We thank J. Khoury for many useful discussions, and H. Bantilan for sharing references on the 3+1 formalism and proofreading the manuscript.
This work is supported in part by Department of Energy Grant DE-FG02-91-ER-40671.

\end{acknowledgments}

\appendix

\section{Generalized Friedmann equations} \label{sec:hypersurface}

\subsection{spatial hypersurfaces and timelike congruences}

Consider a spacetime $(\mathcal{M}, \pmb{g})$ foliated by a family of spacelike hypersurfaces $\{ \Sigma_\tau , \tau \!\!\in\!\! \mathbb{R} \}$, we may construct a coordinate system that is adapted to the foliation as follows \cite{Gourgoulhon:2007ue}. Take $\tau$ to be the natural time coordinate, then the lapse function $\mathcal{N}$ is defined as
\begin{equation}
\mathcal{N} \equiv \big( - (\partial \tau)^2 \, \big)^{-1/2} .
\end{equation}
Let $\pmb{n}$ be the timelike unit vector normal to the constant time hypersurfaces, then by definition $n_\mu = (-\mathcal{N}, \vec{0} \, )$. The spatial coordinates are fixed by choosing a shift vector $\pmb{\beta}$ defined as
\begin{equation}
\pmb{\beta} \equiv \pmb{\partial}_\tau - \mathcal{N} \pmb{n} \; ,
\end{equation}
which implies $n^\mu = (\frac{1}{\mathcal{N}}, -\frac{\beta^i}{\mathcal{N}})$. Under this coordinate system, the metric can be written as
\begin{equation} \label{eq:metric}
ds^2 = - \mathcal{N}^2 d\tau^2 + \gamma_{ij} ( dx^i + \beta^i d\tau ) ( dx^j + \beta^j d\tau ) ,
\end{equation}
or in matrix form,
\begin{equation}
g_{\mu\nu} = \left( \begin{array}{cc} g_{00} & g_{0j} \\ g_{i0} & g_{ij} \end{array} \right) = \left( \begin{array}{cc} - \mathcal{N}^2 + \beta_k \beta^k & \beta_j \\ \beta_i & \gamma_{ij} \end{array} \right) ,
\end{equation}
where $\beta_i = \gamma_{ij} \beta^j$. Accordingly, the inverse metric is given by
\begin{equation}
g^{\mu\nu} = \left( \begin{array}{cc} g^{00} & g^{0j} \\[3pt] g^{i0} & g^{ij} \end{array} \right) = \left( \begin{array}{cc} - \frac{1}{\mathcal{N}^2} & \frac{\beta^j}{\mathcal{N}^2} \\[3pt] \frac{\beta^i}{\mathcal{N}^2} & \gamma^{ij} - \frac{\beta^i \beta^j}{\mathcal{N}^2} \end{array} \right) ,
\end{equation}
where $\gamma^{ij}$ is the inverse of the 3-matrix $\gamma_{k\ell}$.

The intrinsic curvature of the hypersurface $\Sigma$ is given by the Ricci tensor $^{(3)} \! R_{ij}$ associated with the induced spatial metric $\gamma_{ij}$ on the hypersurface. $\gamma_{ij}$ can be pushed forward to form a projection tensor $\pmb{\gamma}$ in $\mathcal{M}$,
\begin{equation}
\gamma_{\mu\nu} \equiv g_{\mu\nu} + n_\mu n_\nu \; .
\end{equation}
The extrinsic curvature of the hypersurface $\Sigma$ is then given by
\begin{equation}
K_{\mu\nu} \equiv \gamma^\kappa_{\;\; \mu} \gamma^\lambda_{\;\; \nu} n_{\kappa;\lambda} \; ,
\end{equation}
where semicolon $;$ denotes the covariant derivative compatible with the metric (\ref{eq:metric}). It can be shown that $K_{\mu\nu}$ is symmetric, $K_{\mu\nu} = K_{\nu\mu}$, and tangent to the hypersurface, $K_{\mu\nu} n^\nu = 0$. The mean curvature of the hypersurface is $1/3$ of the trace $K \equiv K^\mu_{\;\;\mu} = K^i_{\;\;i}$.

The 4-dimensional Riemann tensor $R^\mu_{\;\;\lambda\nu\kappa}$ of the spacetime $\mathcal{M}$ can be related to the intrinsic and extrinsic curvatures of the hypersurface $\Sigma$ through the Gauss-Codazzi relations \cite{Gourgoulhon:2007ue}, especially,
\begin{align}
& \gamma^\mu_{\;\;\alpha} \gamma^\nu_{\;\;\beta} R_{\mu\lambda\nu\kappa} n^\lambda n^\kappa + \gamma^\mu_{\;\;\alpha} \gamma^\nu_{\;\;\beta} R_{\mu\nu} \nonumber \\
& \quad = {}^{(3)}\!R_{\alpha\beta} + K K_{\alpha\beta} - K_{\alpha\mu} K^\mu_{\;\;\beta} \; , \label{eq:Gauss1} \\[4pt]
& 2 R_{\mu\nu} n^\mu n^\nu + R = {}^{(3)}\!R + K^2 - K^{ij} K_{ij} \; , \label{eq:Gauss2} \\[4pt]
& R_{\mu\nu} n^\mu \gamma^\nu_{\;\;\alpha} = - K_{|\alpha} + K^\mu_{\;\;\alpha|\mu} \; , \label{eq:Codazzi1}
\end{align}
where $|$ denotes covariant derivative associated with the induced spatial metric $\pmb{\gamma}$.

Consider the congruence of the integral curves of the timelike normal vector $\pmb{n}$. Such an integral curve can be regarded as the worldline of an Eulerian observer \cite{Gourgoulhon:2007ue}, who is defined to have a 4-velocity equal to $\pmb{n}$. The covariant derivative of the timelike vector $n_\mu$ can be kinematically decomposed as \cite{Hawking:1973uf}
\begin{equation}
n_{\mu;\nu} = \theta_{\mu\nu} + \omega_{\mu\nu} - a_\mu n_\nu \; ,
\end{equation}
where
\begin{equation}
a_\mu \equiv \dot{n}_\mu \equiv n^\nu n_{\mu;\nu}
\end{equation}
is the acceleration of the Eulerian observer, and the expansion tensor $\theta_{\mu\nu}$ and the vorticity tensor $\omega_{\mu\nu}$ are symmetric and antisymmetric respectively. It can be shown ({\it e.g.} \cite{Carroll:2004st} ) that for the unit vector $\pmb{n}$ normal to the hypersurface $\Sigma$, the vorticity tensor $\omega_{\mu\nu}$ vanishes, and the expansion tensor $\theta_{\mu\nu} = n_{\mu;\nu} + a_\mu n_\nu$ is equal to the extrinsic curvature $K_{\mu\nu}$. Decomposing $\theta_{\mu\nu}$ further into the trace and the traceless parts, we have
\begin{equation} \label{eq:decomp}
n_{\mu;\nu} = \tfrac{1}{3} \theta \gamma_{\mu\nu} + \sigma_{\mu\nu} - a_\mu n_\nu \; ,
\end{equation}
where the volume expansion
\begin{equation}
\theta \equiv n^\mu_{\;\; ;\mu}
\end{equation}
is equal to the trace of extrinsic curvature $K$, and the shear $\sigma_{\mu\nu}$ is traceless and symmetric. Both the shear $\sigma_{\mu\nu}$ and the acceleration $a_\mu$ are tangent to the hypersurface, $\sigma_{\mu\nu} n^\nu = a_\nu n^\nu = 0$. For the general metric (\ref{eq:metric}) and the normal vector $n_\mu = (-\mathcal{N}, \vec{0} \, )$, the expansion, shear, and acceleration can be explicitly computed to be
\begin{align}
\theta = & - \frac{1}{\mathcal{N}} \beta^k_{\; ,k} + \frac{1}{2 \mathcal{N}} \gamma^{ij} \gamma_{ij}^{\;\prime} - \frac{1}{2 \mathcal{N}} \gamma^{ij} \gamma_{ij,k} \beta^k \; , \label{eq:localtheta} \\[4pt]
\sigma^{ij} = & - \frac{1}{2 \mathcal{N}} \gamma^{ij \, \prime} - \frac{1}{\mathcal{N}} \gamma^{k(i} \beta^{j)}_{\; ,k} + \frac{1}{2 \mathcal{N}} \gamma^{ij}_{\;\; ,k} \beta^k \label{eq:localsigmaij} \\
& + \frac{1}{6 \mathcal{N}} \gamma^{ij} \big( - \gamma^{k\ell} \gamma_{k\ell}^{\;\prime} + 2 \beta^k_{\; ,k} + \gamma^{k\ell} \gamma_{k\ell ,m} \beta^m \big) \; , \nonumber \\[4pt]
a^i = &\ \frac{1}{\mathcal{N}} \gamma^{ij} \mathcal{N}_{,j} \; . \label{eq:locala}
\end{align}

Using the kinematic decomposition, the Gauss-Codazzi relations (\ref{eq:Gauss1}, \ref{eq:Gauss2}, \ref{eq:Codazzi1}) can be written as
\begin{align}
& \gamma^\mu_{\;\;\alpha} \gamma^\nu_{\;\;\beta} R_{\mu\lambda\nu\kappa} n^\lambda n^\kappa + \gamma^\mu_{\;\;\alpha} \gamma^\nu_{\;\;\beta} R_{\mu\nu} \label{eq:Gauss1v} \nonumber \\
& \quad = {}^{(3)}\!R_{\alpha\beta} + \tfrac{2}{9} \theta^2 \gamma_{\alpha\beta} + \tfrac{1}{3} \theta \sigma_{\alpha\beta} - \sigma_{\alpha\mu} \sigma^\mu_{\;\;\beta} \; , \\[4pt]
& 2 R_{\mu\nu} n^\mu n^\nu + R = {}^{(3)}\!R + \tfrac{2}{3} \theta^2 - 2 \sigma^2 \; , \label{eq:Gauss2v} \\[4pt]
& R_{\mu\nu} n^\mu \gamma^\nu_{\;\;\alpha} = - \tfrac{2}{3} \theta_{|\alpha} + \sigma^\mu_{\;\;\alpha|\mu} \; , \label{eq:Codazzi1v}
\end{align}
where $\sigma^2 \equiv \frac{1}{2} \sigma^{ij} \sigma_{ij}$ is the squared magnitude of the shear \cite{Hawking:1973uf, Bardeen:1980kt, Kodama:1985bj}.

To study the kinematic evolution of the timelike congruence, let us first introduce the Fermi derivative $\frac{D_F}{ds}$ \cite{Hawking:1973uf} that propagates a timelike vector $\pmb{V}$ along its integral curve, $\frac{D_F}{ds} \pmb{V} = 0$. The Fermi derivative of a vector field $\pmb{X}$ with respect to $\pmb{V}$ is defined as
\begin{equation}
\frac{D_F}{ds} X^\mu \equiv \dot{X}^\mu - V^\mu \dot{V}_\nu X^\nu + \dot{V}^\mu V_\nu X^\nu ,
\end{equation}
while for a covariant vector $\underline{\pmb{\omega}}$ it is
\begin{equation}
\frac{D_F}{ds} \omega_\mu \equiv \dot{\omega}_\mu - V_\mu \dot{V}^\nu \omega_\nu + \dot{V}_\mu V^\nu \omega_\nu ,
\end{equation}
and similarly for tensors. It reduces to the covariant derivative $\frac{D}{ds} X^\mu \equiv \dot{X}^\mu \equiv V^\nu X^\mu_{\;\; ;\nu}$ if the integral curve of $\pmb{V}$ is a geodesic, {\it i.e.} when $\dot{V}^\mu = 0$.

The Fermi derivative of the expansion tensor $\theta_{\mu\nu}$ with respect to $\pmb{n}$ is given by \cite{Hawking:1973uf}
\begin{align}
\frac{D_F}{ds} \theta_{\mu\nu} = &- R_{\mu\lambda\nu\kappa} n^\lambda n^\kappa - \theta_{\mu\lambda} \theta^\lambda_{\;\;\nu} - 2 n_{(\mu} \theta_{\nu)\lambda} a^\lambda \nonumber \\
&+ \gamma^\lambda_{\;\;\mu} \gamma^\kappa_{\;\;\nu} a_{(\lambda;\kappa)} + a_\mu a_\nu \; .
\end{align}
The trace of this equation gives the Raychaudhuri equation,
\begin{equation} \label{eq:Raychaudhuri}
\dot{\theta} = - R_{\lambda\kappa} n^\lambda n^\kappa - \tfrac{1}{3} \theta^2 - 2 \sigma^2 + a^\mu_{\;\; ;\mu} .
\end{equation}
And the traceless part of the equation gives
\begin{align}
\frac{D_F}{ds} \sigma_{\mu\nu} = &- R_{\mu\lambda\nu\kappa} n^\lambda n^\kappa - \tfrac{2}{3} \theta \sigma_{\mu\nu} - \sigma_{\mu\lambda} \sigma^\lambda_{\;\;\nu} + \gamma^\lambda_{\;\;\mu} \gamma^\kappa_{\;\;\nu} a_{(\lambda;\kappa)} \nonumber \\
&+ a_\mu a_\nu + \tfrac{1}{3} \gamma_{\mu\nu} \big( R_{\lambda\kappa} n^\lambda n^\kappa + 2 \sigma^2 - a^\lambda_{\;\; ;\lambda} \big) .
\end{align}
Multiplying by $\gamma^{\mu i} \gamma^\nu_{\;\; j}$ and using Eqs.~(\ref{eq:Gauss1v}, \ref{eq:Gauss2v}), we find
\begin{align} \label{eq:DFsigmaij}
\frac{D_F}{ds} \sigma^i_{\ j} = &\ \gamma^{\mu i} \gamma^\nu_{\;\; j} R_{\mu\nu} - {}^{(3)}\!R^i_{\ j} - \theta \sigma^i_{\ j} - n^i \sigma_{jk} a^k \nonumber \\
& - \tfrac{1}{3} \theta n^i a_j + a^i_{\ ;j} + a^i a_j + \dot{a}^i n_j \nonumber \\
& - \tfrac{1}{3} \gamma^i_{\ j} ( R + R_{\mu\nu} n^\mu n^\nu - {}^{(3)}\!R + a^\mu_{\;\; ;\mu} ) .
\end{align}

\subsection{local Friedmann equations}

The dynamics of the timelike congruence is determined by the Einstein equation relating the spacetime curvature and the matter stress-energy tensor. Consider the matter source to be a perfect fluid with stress-energy tensor
\begin{equation} \label{eq:perfectfluid}
T_{\mu\nu} = (\rho + P) u_\mu u_\nu + P g_{\mu\nu} \; ,
\end{equation}
where $\rho$ and $P$ are the rest energy density and pressure, and $\pmb{u}$ is the 4-velocity of the fluid. The fluid velocity relative to the Eulerian observer is
\begin{equation}
\pmb{U} = \frac{\pmb{u}}{\Gamma} - \pmb{n} \; ,
\end{equation}
where the Lorentz factor $\Gamma$ is given by
\begin{equation}
\Gamma = - n^\mu u_\mu = (1 - U^\mu U_\mu)^{-1/2} .
\end{equation}
Accordingly \cite{Gourgoulhon:2007ue}, the energy density as measured by the Eulerian observer is
\begin{equation}
E = T_{\mu\nu} n^\mu n^\nu = \Gamma^2 (\rho + P) - P \; ,
\end{equation}
the momentum density as measured by the Eulerian observer is
\begin{equation}
p_\alpha = - T_{\mu\nu} n^\mu \gamma^\nu_{\;\;\alpha} = (E + P) U_\alpha \; ,
\end{equation}
and the stress tensor with respect to the Eulerian observer is
\begin{equation}
S_{\mu\nu} = \gamma^\lambda_{\;\;\mu} \gamma^\kappa_{\;\;\nu} T_{\lambda\kappa} = (E + P) U_\mu U_\nu + P \gamma_{\mu\nu} \; .
\end{equation}
The trace of the stress-energy tensor is given by $T \equiv T^\mu_{\;\;\mu} = - \rho + 3 P$, while the trace of the relative stress tensor is $S \equiv S^\mu_{\;\;\mu} = T + E$. The Einstein equation can be written as
\begin{equation}
R_{\mu\nu} = T_{\mu\nu} - \tfrac{1}{2} g_{\mu\nu} T \; ,
\end{equation}
the trace of which gives
\begin{equation}
R = - T = \rho - 3 P \; .
\end{equation}
Projecting the Einstein equation along the normal vector $\pmb{n}$, we find
\begin{equation}
R_{\mu\nu} n^\mu n^\nu = E + \tfrac{1}{2} ( - \rho + 3 P ) \; ,
\end{equation}
whereas the mixed projection with $\pmb{n}$ and $\pmb{\gamma}$ gives
\begin{equation}
R_{\mu\nu} n^\mu \gamma^\nu_{\;\;\alpha} = - p_\alpha = - (E + P) U_\alpha \; ,
\end{equation}
and the full projection onto the hypersurface $\Sigma$ gives
\begin{align}
\gamma^{\mu i} \gamma^\nu_{\;\; j} R_{\mu\nu} &= S^i_{\ j} - \tfrac{1}{2} \gamma^i_{\ j} ( - \rho + 3 P ) \nonumber \\
&= (E + P) U^i U_j + \tfrac{1}{2} \gamma^i_{\ j} \rho \; .
\end{align}

Finally, we can use the above equations to write Eqs.~(\ref{eq:Gauss2v}), (\ref{eq:Raychaudhuri}), (\ref{eq:Codazzi1v}) and (\ref{eq:DFsigmaij}) as
\begin{align}
(\tfrac{1}{3} \theta)^2 &= \tfrac{1}{3} \big( E - \tfrac{1}{2} {}^{(3)}\!R + \sigma^2 \big) \; , \label{eq:localFriedmann1} \\[4pt]
\tfrac{1}{3} \dot{\theta} &= - \tfrac{1}{2} \big( \tfrac{4 E - \rho}{3} + P \big) + \tfrac{1}{6} {}^{(3)}\!R - \sigma^2 + \tfrac{1}{3} a^\mu_{\;\; ;\mu} \; , \label{eq:localFriedmann2} \\[4pt]
\tfrac{1}{3} \theta_{|i} &= \tfrac{1}{2} \big( E + P \big) U_i + \tfrac{1}{2} \sigma^j_{\ i|j} \; , \label{eq:localFriedmann3}
\end{align}
and
\begin{align} \label{eq:localDFsigmaij}
\frac{D_F}{ds} \sigma^i_{\ j} = &\ (E + P) U^i U_j - {}^{(3)}\!R^i_{\ j} - \theta \sigma^i_{\ j} \nonumber \\
& - n^i \sigma_{jk} a^k - \tfrac{1}{3} \theta n^i a_j + a^i_{\ ;j} + a^i a_j + \dot{a}^i n_j \nonumber \\
& - \tfrac{1}{3} \delta^i_{\ j} ( E - \rho - {}^{(3)}\!R + a^\mu_{\;\; ;\mu} ) .
\end{align}
Eqs.~(\ref{eq:localFriedmann1}) and (\ref{eq:localFriedmann3}) are equivalent to the $(^0_0)$ and $(^0_i)$ components of the Einstein equation; Eq.~(\ref{eq:localFriedmann2}) corresponds to the trace of the $(^i_j)$ components, whereas Eq.~(\ref{eq:localDFsigmaij}) corresponds to the traceless part.

These equations closely resemble the Friedmann equations for a homogeneous universe. The expansion $\theta$ can be considered as 3 times the local Hubble parameter \cite{Lyth:2004gb}, then Eqs.~(\ref{eq:localFriedmann1}) and (\ref{eq:localFriedmann2}) are the \emph{local Friedmann equations}. Note that according to Eq.~(\ref{eq:localFriedmann3}) the local Hubble parameter may vary from point to point on the constant time hypersurface. In the homogeneous case, the fluid stress-energy tensor $T_{\mu\nu}$ is diagonal; the expansion $\theta$ and the shear $\sigma^i_{\; j}$ do not depend on spatial coordinates, while the acceleration $a^\mu$ vanishes. Accordingly, Eqs.~(\ref{eq:localFriedmann1}, \ref{eq:localFriedmann2}) reduce to the Friedmann equations,
\begin{align}
H^2 &= \tfrac{1}{3} \big( \rho - \tfrac{1}{2} {}^{(3)}\!R + \sigma^2 \big) \; , \label{eq:Friedmann1homo} \\[4pt]
\dot{H} &= - \tfrac{1}{2} \big( \rho + P \big) + \tfrac{1}{6} {}^{(3)}\!R - \sigma^2 \; . \label{eq:Friedmann2homo}
\end{align}
Eq.~(\ref{eq:localFriedmann3}) vanishes identically, whereas Eq.~(\ref{eq:localDFsigmaij}) simplifies to
\begin{equation}
\dot{\sigma}^i_{\; j} = - 3 H \sigma^i_{\; j} - {}^{(3)}\!R^i_{\; j}  + \tfrac{1}{3} \delta^i_{\; j} {}^{(3)}\!R \; . \label{eq:Friedmann4homo}
\end{equation}

\subsection{gauge choices}

We may choose certain gauge for the coordinates where equations (\ref{eq:localtheta}$\sim$\ref{eq:locala}) and (\ref{eq:localFriedmann1}$\sim$\ref{eq:localDFsigmaij}) simplify. First let
\begin{equation} \label{eq:detgamma}
\gamma_{ij} \equiv a^2 \; \tilde{\gamma}_{ij} \; ,
\end{equation}
where the scale factor $a$ is such that $\det \tilde{\gamma} = 1$. The conformal Hubble parameter is therefore $\mathcal{H} \equiv \frac{a'}{a}$. Note that here the scale factor $a$ may depend on the spatial coordinates, and so does $\mathcal{H}$. Then the expansion, shear, and acceleration become
\begin{align}
\theta = &\ \frac{3}{\mathcal{N}} \mathcal{H} - \frac{3}{\mathcal{N}} \frac{a_{,k}}{a} \beta^k - \frac{1}{\mathcal{N}} \beta^k_{\; ,k} \; , \\[4pt]
\sigma^{ij} = & - \frac{1}{2 \mathcal{N} a^2} \tilde{\gamma}^{ij \, \prime} - \frac{1}{\mathcal{N} a^2} \tilde{\gamma}^{k(i} \beta^{j)}_{\; ,k} + \frac{1}{2 \mathcal{N} a^2} \tilde{\gamma}^{ij}_{\;\; ,k} \beta^k \nonumber \\
& + \frac{1}{3 \mathcal{N} a^2} \tilde{\gamma}^{ij} \beta^k_{\; ,k} \; , \\[4pt]
a^i = &\ \frac{1}{\mathcal{N} a^2} \tilde{\gamma}^{ij} \mathcal{N}_{,j} \; .
\end{align}
To specify a gauge, the time slicing can be fixed by either choosing a foliation condition for the whole spacetime, or by starting from an initial Cauchy surface and choosing a lapse function $\mathcal{N}$. Then the spatial coordinates can be fixed by choosing a shift vector $\beta^i$.

The \textbf{comoving gauge} is defined such that the foliation $\{ \Sigma_\tau \}$ is everywhere orthogonal to the fluid velocity $\pmb{u}$, {\it i.e.} the Eulerian observer is comoving with the fluid, so that $\pmb{n} = \pmb{u}$. This gauge exists when the fluid does not have vorticity, which is true for a scalar field. In this gauge the relative velocity $\pmb{U}$ vanishes, so the Eulerian observer simply measures the comoving energy density and stress tensor,
\begin{equation}
E = \rho \; , \quad S_{\mu\nu} = P \gamma_{\mu\nu} \; .
\end{equation}
We may further fix the spatial gauge by choosing $\beta^i = 0$, so that the expansion and the shear simplify to
\begin{equation} \label{eq:theta,sigma}
\theta = \frac{3}{\mathcal{N}} \mathcal{H} \; , \quad \sigma^i_{\; j} = \frac{1}{2 \mathcal{N}} \tilde{\gamma}^{ik} \tilde{\gamma}_{kj}^{\;\prime} \; .
\end{equation}
Here the expansion $\theta$ may depend on spatial coordinates, hence the local Hubble parameter is not uniform on the constant time hypersurface. Also, since the fluid is in general not free streaming, the acceleration $a^\mu$ does not vanish. The local Friedmann equations become
\begin{align}
(\tfrac{1}{3} \theta)^2 &= \tfrac{1}{3} \big( \rho - \tfrac{1}{2} {}^{(3)}\!R + \sigma^2 \big) \; , \label{eq:comovingH2} \\[4pt]
\tfrac{1}{3} \dot{\theta} &= - \tfrac{1}{2} \big( \rho + P \big) + \tfrac{1}{6} {}^{(3)}\!R - \sigma^2 + \tfrac{1}{3} a^\mu_{\;\; ;\mu} \; , \label{eq:comovingHdot} \\[4pt]
\tfrac{1}{3} \theta_{|i} &= \tfrac{1}{2} \sigma^j_{\ i|j} \; , \label{eq:comovingH_i}
\end{align}
and the shear $\sigma^i_{\ j}$ evolves according to
\begin{align}
\frac{D_F}{ds} \sigma^i_{\ j} = & - \theta \sigma^i_{\ j} - {}^{(3)}\!R^i_{\ j}  + \tfrac{1}{3} \delta^i_{\ j} ( {}^{(3)}\!R - a^\mu_{\;\; ;\mu} ) \label{eq:comovingDFsigmaij} \\
& - u^i \sigma_{jk} a^k - \tfrac{1}{3} \theta u^i a_j + a^i_{\ ;j} + a^i a_j + \dot{a}^i u_j \; . \nonumber
\end{align}

The \textbf{synchronous gauge} is defined such that $\mathcal{N} = 1$ and $\beta^i = 0$, which fixes the coordinates given the choice of an initial Cauchy surface and spatial coordinates therein. In this gauge the acceleration vanishes according to Eq.~(\ref{eq:locala}), so the integral curve of $\pmb{n}$, or the worldline of the Eulerian observer, is a geodesic. Then Eq.~(\ref{eq:localFriedmann2}) becomes
\begin{equation}
\tfrac{1}{3} \dot{\theta} = - \tfrac{1}{2} \big( \tfrac{4 E - \rho}{3} + P \big) + \tfrac{1}{6} {}^{(3)}\!R - \sigma^2 \; ,
\end{equation}
and Eq.~(\ref{eq:localDFsigmaij}) simplifies to
\begin{equation}
\dot{\sigma}^i_{\; j} = - \theta \sigma^i_{\; j} + (E + P) U^i U_j - {}^{(3)}\!R^i_{\ j} - \tfrac{1}{3} \delta^i_{\ j} ( E - \rho - {}^{(3)}\!R ) ,
\end{equation}
where the Fermi derivative reduces to the covariant derivative.

Another useful gauge is the \textbf{constant mean curvature gauge} in which the foliation is such that the mean curvature $\theta/3$ is constant on each time slice. In other words, the Hubble parameter $H$ is uniform on each constant time hypersurface, and does not depend on the spatial coordinates. In this gauge, the Friedmann equations (\ref{eq:localFriedmann1}, \ref{eq:localFriedmann2}) become
\begin{align}
H^2 &= \tfrac{1}{3} \big( E - \tfrac{1}{2} {}^{(3)}\!R + \sigma^2 \big) \; , \label{eq:CMCH2} \\[4pt]
\dot{H} &= - \tfrac{1}{2} \big( \tfrac{4 E - \rho}{3} + P \big) + \tfrac{1}{6} {}^{(3)}\!R - \sigma^2 + \frac{1}{3} a^\mu_{\;\; ;\mu} \; , \label{eq:CMCHdot}
\end{align}
and Eq.~(\ref{eq:localFriedmann3}) becomes a constraint on the shear,
\begin{equation} \label{eq:sigma-constr}
\sigma^j_{\ i|j} = - ( E + P ) U_i .
\end{equation}

\section{Linear perturbations} \label{sec:perturbation}

On the perturbative level, the general metric (\ref{eq:metric}) can be expanded about the flat FRW metric as, to linear order,
\begin{align} \label{eq:metricpert}
ds^2 = &\ a(\tau)^2 \Big[ - ( 1 + 2 A) d\tau^2 + 2 ( B_{,i} + S_i ) d\tau dx^i \\
& + \big( ( 1 - 2 \psi ) \delta_{ij} + 2 E_{,ij} + 2 F_{(i,j)} + 2 h_{ij} \big) dx^i dx^j \Big] , \nonumber
\end{align}
where $A$, $B$, $\psi$ and $E$ represent the scalar perturbations; $S_i$ and $F_i$, with $S^i_{\; ,i} = F^i_{\; ,i} = 0$, represent the vector perturbations; and $h_{ij}$, with $h^i_{\; i} = h^i_{\; j,i} = 0$, represent the tensor perturbations \cite{Mukhanov:1990me}. Comparing to (\ref{eq:metric}), we identify the lapse, shift and the spatial 3-metric to be, also to linear order,
\begin{align}
\mathcal{N} &= a \, ( 1 + A ) , \\[4pt]
\beta_i &= a^2 ( B_{,i} + S_i ) , \\[4pt]
\gamma_{ij} &= a^2 \big( ( 1 - 2 \psi ) \delta_{ij} + 2 E_{,ij} + 2 F_{(i,j)} + 2 h_{ij} \big) . \label{eq:gamma_pert}
\end{align}
Accordingly, the constant time hypersurface has a timelike normal vector $n_\mu = ( -a (1+A), \vec{0} \, )$. The intrinsic curvature of the hypersurface is given by \cite{Wands:2000dp}
\begin{equation} \label{eq:3R_pert}
{}^{(3)} \! R = \frac{4}{a^2} \nabla^2 \psi \; ,
\end{equation}
whereas the trace of the extrinsic curvature $K$, or equally the expansion $\theta$, is given by
\begin{equation} \label{eq:theta_pert}
\theta = \frac{1}{a} \big[ 3 \mathcal{H} ( 1 - A ) - 3 \psi^\prime + \nabla^2 (E^\prime - B) \big] .
\end{equation}
Note that the expansion only involves the scalar perturbations. The shear of the constant time hypersurface is given by \cite{Kodama:1985bj}
\begin{align} \label{eq:sigma_pert}
\sigma_{ij} = & \; a \big( (E^\prime_{,ij} \! - \! B_{,ij}) - \tfrac{1}{3} \delta_{ij} \nabla^2 (E^\prime \! - \! B) \big) + a \big( F_{(i,j)}^\prime \! - \! S_{(i,j)} \big) \nonumber \\
\equiv & \; a \big( \sigma^\text{S}_{,ij} - \tfrac{1}{3} \delta_{ij} \nabla^2 \sigma^\text{S} \big) + a \, \sigma^\text{V}_{(i,j)} 
\end{align}
where the scalar shear perturbation is defined as
\begin{equation}
\sigma^\text{S} \equiv E^\prime - B \; ,
\end{equation}
and the vector shear perturbation is
\begin{equation}
\sigma^\text{V}_i \equiv F_i^\prime - S_i \ .
\end{equation}

For the matter source, consider a scalar field $\phi$ with Lagrangian $\mathcal{L} = P(\phi, X)$, where $X = - \frac{1}{2} (\partial \phi)^2$. The corresponding stress-energy tensor is
\begin{equation}
T_{\mu\nu} = P_{,X} \, \partial_\mu \phi \, \partial_\nu \phi + P \, g_{\mu\nu} \; ,
\end{equation}
which takes the form of a perfect fluid (\ref{eq:perfectfluid}) with pressure $P$, energy density $\rho = 2 X P_{,X} - P \,$, and velocity $u_\mu = \partial_\mu \phi / \sqrt{2X} \,$. In accordance with the homogeneous FRW background metric, the scalar field is assumed to be homogeneous on the background level as well. Therefore the background values $\phi$, $X$, and hence $\rho$, $P$, are all functions of time $\tau$ only, and $u_\mu = (-a, \vec{0}\, )$. Since the shear vanishes at this level, the background equations of motion are fully captured by the Friedmann equations,
\begin{align}
H^2 &= \tfrac{1}{3} \rho = \tfrac{1}{3} ( 2 X P_{,X} - P ) \; , \label{eq:Friedmann1-bgd} \\[4pt]
\dot{H} &= - \tfrac{1}{2} ( \rho + P ) = - X P_{,X} \; . \label{eq:Friedmann2-bgd}
\end{align}

Then let us consider a small perturbation $\delta\phi (\tau, \vec{x} \,)$ about the background value $\phi (\tau)$, which also creates a perturbation about $X (\tau)$, $\delta X = 2 X ( - A + \frac{\delta\phi^\prime}{\phi^\prime} )$. Accordingly, the perturbations in velocity, energy density, and pressure are given by, to linear order,
\begin{align}
& \delta u_0 = - a A \; , \quad \delta u_i = a \, \delta u_{,i} \; , \quad \delta u \equiv \Big( \frac{\delta\phi}{- \phi^\prime} \Big) \; , \label{eq:delta_u} \\
& \delta \rho = - \frac{2}{a^2} ( \mathcal{H}^\prime \! - \! \mathcal{H}^2 ) \Big[ \, \frac{1}{c_s^2} \big( \delta u^\prime + \mathcal{H} \delta u - A \big) - 3 \mathcal{H} \delta u \Big] , \label{eq:delta_rho} \\
& \delta P = - \frac{2}{a^2} ( \mathcal{H}^\prime \! - \! \mathcal{H}^2 ) \Big[ \, \delta u^\prime + \Big( 2 \mathcal{H} + \frac{( \mathcal{H}^\prime \! - \! \mathcal{H}^2 )^\prime}{( \mathcal{H}^\prime \! - \! \mathcal{H}^2 )} \Big) \delta u - A \Big] , \label{eq:delta_P}
\end{align}
where the speed of sound is $c_s^2 = \frac{P_{,X}}{\rho_{,X}}$, which only needs be kept to zeroth order. Here we have assumed $\phi^\prime < 0$ to agree with our model in the text.

The equations of motion for the linear perturbations are given by the perturbed Einstein equation, $\delta G^\mu_{\;\nu} = \delta T^\mu_{\;\nu}$. (Alternatively, one may impose the conservation of energy and momentum \cite{Wands:2000dp}, plus the equation of motion for the scalar field.) At linear order, the scalar, vector, and tensor perturbations evolve independently and can be treated separately. In the following we discuss the scalar and vector perturbations.

\subsection{scalar perturbations}

The equations for the scalar perturbations are
\begin{align}
& \frac{2}{a^2} \Big[ 3 \mathcal{H} ( \psi^\prime \! + \! \mathcal{H} A ) - \nabla^2 \big( \psi + \mathcal{H} ( E^\prime \! - \! B ) \big) \Big] = - \delta \rho , \label{eq:G00} \\[4pt]
& \frac{2}{a^2} \Big[ - ( \psi^\prime \! + \! \mathcal{H} A ) \Big] = ( \rho + P ) \delta u \; , \label{eq:G0i} \\[4pt]
& \frac{2}{a^2} \Big[ \psi - A + ( E^\prime \! - \! B )^\prime + 2 \mathcal{H} ( E^\prime \! - \! B ) \Big] = 0 \; , \label{eq:Gij} \\[4pt]
& \frac{2}{a^2} \Big[ ( \psi^\prime \! + \! \mathcal{H} A )^\prime + 2 \mathcal{H} ( \psi^\prime \! + \! \mathcal{H} A ) + ( \mathcal{H}^\prime \! - \! \mathcal{H}^2 ) A \Big] = \delta P \; , \label{eq:Gii}
\end{align}
where the matter perturbations are given in terms of the scalar field perturbation $\delta\phi$ through Eqs.~(\ref{eq:delta_u}, \ref{eq:delta_rho}, \ref{eq:delta_P}). These equations correspond to the scalar modes of the $(^0_{\;0})$, $(^0_{\;i})$, traceless $(^i_{\;j})$, and the trace $(^i_{\;i})$ components of the Einstein equation. There are 4 degrees of freedom in these equations, namely $A$, $\psi$, $\sigma^\text{S} \! = \! E^\prime \! - \! B$, and $\delta\phi$. However, as we shall see below, these variables are not independent since there exists one gauge redundancy. Therefore, after gauge fixing, we are left with 3 physical degrees of freedom. Accordingly, given the matter perturbations in the form (\ref{eq:delta_u}) $\sim$ (\ref{eq:delta_P}), one of the equations in (\ref{eq:G00}) $\sim$ (\ref{eq:Gii}) is redundant.

Consider the infinitesimal coordinate transformation
\begin{equation} \label{eq:coordtransS}
x^\mu \rightarrow x^\mu + \xi^\mu \; ,
\end{equation}
where for the scalar mode, $\xi_i$ can be written as $\xi_i \equiv \xi_{,i}$. From the transformation rules of the metric tensor $g_{\mu\nu}$ and the scalar $\phi$, one finds that the perturbative quantities transform as
\begin{align}
A &\rightarrow A - \tfrac{1}{a} ( a \, \xi^0 )^\prime \; , \\[3pt]
\psi &\rightarrow \psi + \mathcal{H} \, \xi^0 \; , \label{eq:psi_transf} \\[3pt]
\sigma^\text{S} &\rightarrow \sigma^\text{S} - \xi^0 \; , \label{eq:sigma_transf} \\[3pt]
\delta\phi &\rightarrow \delta\phi - \phi^\prime \, \xi^0 \; .
\end{align}
Thus they can be unambiguously defined provided a specified choice of time slicing, $\xi^0$. Some commonly used gauges are \cite{Kodama:1985bj}: the synchronous gauge in which one sets $A = 0$; the flat gauge in which $\psi = 0$; the longitudinal gauge in which $\sigma^\text{S} = 0$; the comoving gauge in which $\delta u = 0$, or equivalently $\delta\phi = 0$ in our case; the uniform density gauge in which $\delta \rho = 0$; and the uniform Hubble gauge in which $\theta = 3 \mathcal{H} / a$, as detailed below. To completely fix the gauge, one also needs to specify a choice of spatial coordinates, $\xi$. Since $B$ and $E$ transform as
\begin{align}
B & \rightarrow B + \xi^0 - \xi^\prime \; , \\[3pt]
E & \rightarrow E - \xi \; ,
\end{align}
one commonly chooses to set $B = 0$. \\

\noindent \textbf{Comoving gauge: $\delta u = \delta\phi = 0$.} This gauge is defined only for $\phi^\prime \neq 0$. The physical degrees of freedom in this gauge can be represented by the gauge-invariant quantities
\begin{align}
A_\text{c} &\equiv A + \frac{1}{a} ( a \, \delta u )^\prime = A + \frac{1}{a} \Big( a \, \frac{\delta\phi}{- \phi^\prime} \Big)^\prime \; , \\
\mathcal{R} &\equiv \psi - \mathcal{H} \, \delta u = \psi - \mathcal{H} \Big( \frac{\delta\phi}{- \phi^\prime} \Big) \; , \label{eq:psicom} \\
\sigma_\text{c} &\equiv \sigma^\text{S} + \delta u = \sigma^\text{S} + \Big( \frac{\delta\phi}{- \phi^\prime} \Big) \; . \label{eq:sigmacom}
\end{align}
The equations of motion are the simplest in this gauge,
\begin{align}
& - \nabla^2 \mathcal{R} - \mathcal{H} \, \nabla^2 \sigma_\text{c} = - ( \mathcal{H}^\prime \! - \! \mathcal{H}^2 ) \tfrac{1}{c_s^2} \, A_\text{c} \; , \label{eq:G00com} \\
& \mathcal{R}^\prime + \mathcal{H} A_\text{c} = 0 \; , \label{eq:G0icom} \\[3pt]
& \mathcal{R} - A_\text{c} + \sigma_\text{c}^\prime + 2 \mathcal{H} \sigma_\text{c} = 0 \; . \label{eq:Gijcom}
\end{align}
Eliminating $A_\text{c}$ and $\sigma_\text{c}$, one obtains a simple equation for $\mathcal{R}$ alone,
\begin{equation} \label{eq:Rcom}
\mathcal{R}^{\prime \prime} + 2 \, \frac{z^\prime}{z} \mathcal{R}^{\prime} - c_s^2 \, \nabla^2 \mathcal{R} = 0 ,
\end{equation}
where $z = a \sqrt{{-2 ( \mathcal{H}^\prime \! - \! \mathcal{H}^2 )} / {c_s^2 \mathcal{H}^2}}$ . \\

\noindent \textbf{Longitudinal gauge: $\sigma^\text{S} = 0$.} Perturbations in this gauge \cite{Mukhanov:1990me} are represented by the gauge-invariant quantities
\begin{align}
\Phi &\equiv A - \frac{1}{a} ( a \, \sigma^\text{S} )^\prime \; , \\
\Psi &\equiv \psi + \mathcal{H} \, \sigma^\text{S} \; , \label{eq:psiNewt} \\[3pt]
\overline{\delta\phi} &\equiv \delta\phi - \phi^\prime \, \sigma^\text{S} \; . \label{eq:delta_phiNewt}
\end{align}
$\Phi$ is referred to as the Newtonian potential, which is equal to the longitudinal curvature perturbation $\Psi$ as a result of Eq.~(\ref{eq:Gij}), $\Phi - \Psi = 0$. Hence it suffices to have two more equations of motion,
\begin{align}
& 3 \mathcal{H} \, ( \Psi^\prime + \mathcal{H} \, \Phi ) - \nabla^2 \Psi = - \frac{a^2}{2} \, \overline{\delta \rho} \; , \label{eq:G00Newt} \\
& \Psi^\prime + \mathcal{H} \, \Phi = ( \mathcal{H}^\prime \! - \! \mathcal{H}^2 ) \, \overline{\delta u} \; , \label{eq:G0iNewt}
\end{align}
where $\overline{\delta u} = ( \frac{\overline{\delta\phi}}{- \phi^\prime} )$, and
\begin{equation}
\overline{\delta \rho} = - \frac{2}{a^2} ( \mathcal{H}^\prime \! - \! \mathcal{H}^2 ) \Big[ \, \frac{1}{c_s^2} \big( \overline{\delta u} {}^\prime + \mathcal{H} \, \overline{\delta u} - \Phi \big) - 3 \mathcal{H} \, \overline{\delta u} \Big] . \label{eq:delta_rhoNewt}
\end{equation}
It is easy to relate the perturbations in the longitudinal gauge to that in the comoving gauge. In particular, by Eq.~(\ref{eq:delta_phiNewt}), we have
\begin{equation} \label{eq:deltauNewt}
\overline{\delta u} = \Big( \frac{\delta\phi}{- \phi^\prime} \Big) + \sigma^\text{S} = \sigma_\text{c} \; ,
\end{equation}
and similarly, by Eqs.~(\ref{eq:psiNewt}), (\ref{eq:G00com}) and (\ref{eq:G0icom}),
\begin{equation}
\Psi = \mathcal{R} + \mathcal{H} \, \sigma_\text{c} = - \frac{\mathcal{H}^\prime \! - \! \mathcal{H}^2}{\mathcal{H} \, c_s^2} \, \nabla^{-2} \, \mathcal{R}^\prime \; .
\end{equation}

\noindent \textbf{Synchronous gauge: $A = 0$.} This gauge has a residual degree of freedom, $\xi^0 = C(\vec{x}) / a$, where $C(\vec{x})$ is an arbitrary function of spatial coordinates. We can construct the following quantities that represent the perturbations in this gauge,
\begin{align}
\psi_\text{s} &\equiv \psi + \frac{\mathcal{H}}{a} \int^\tau a A d\tau' \; , \label{eq:psisync} \\
\sigma_\text{s} &\equiv \sigma^\text{S} - \frac{1}{a} \int^\tau a A d\tau' \; , \label{eq:sigmasync} \\
\delta\phi_\text{s} &\equiv \delta\phi - \frac{\phi^\prime}{a} \int^\tau a A d\tau' \; ,
\end{align}
which are invariant up to the residual gauge freedom that can be absorbed into the constant of integration. The equations of motion for these physical degrees of freedom in this gauge can be expressed as
\begin{align}
& 3 \mathcal{H} \psi_\text{s}^\prime - \nabla^2 \psi_\text{s} - \mathcal{H} \, \nabla^2 \sigma_\text{s} = - \frac{a^2}{2} \delta \rho_\text{s} \; , \label{eq:G00sync} \\
& \psi_\text{s}^\prime = ( \mathcal{H}^\prime \! - \! \mathcal{H}^2 ) \delta u_\text{s} \; , \label{eq:G0isync} \\[4pt]
& \psi_\text{s} + \sigma_\text{s}^\prime + 2 \mathcal{H} \sigma_\text{s} = 0 \; , \label{eq:Gijsync}
\end{align}
where $\delta u_\text{s} = ( \frac{\delta\phi_\text{s}}{- \phi^\prime} )$, and
\begin{equation}
\delta \rho_\text{s} = - \frac{2}{a^2} ( \mathcal{H}^\prime \! - \! \mathcal{H}^2 ) \Big[ \, \frac{1}{c_s^2} \big( \delta u_\text{s}^\prime + \mathcal{H} \delta u_\text{s} \big) - 3 \mathcal{H} \delta u_\text{s} \Big] . \label{eq:delta_rhosync}
\end{equation}
By Eqs.~(\ref{eq:psisync}) and (\ref{eq:G0icom}), the synchronous curvature perturbation $\psi_\text{s}$ can be related to the comoving curvature perturbation $\mathcal{R}$ through
\begin{equation} \label{eq:R-to-psi_s}
\psi_\text{s} = \mathcal{R} + \frac{\mathcal{H}}{a} \int^\tau a A_\text{c} d\tau' = \mathcal{R} - \frac{\mathcal{H}}{a} \int^\tau \frac{a}{\mathcal{H}} \, \mathcal{R}^\prime d\tau' \; .
\end{equation}
Similarly, by Eqs.~(\ref{eq:sigmasync}) and (\ref{eq:G0icom}), the synchronous shear perturbation $\sigma_\text{s}$ can be expressed in the comoving gauge as
\begin{equation} \label{eq:sigma_c-to-sigma_s}
\sigma_\text{s} = \sigma_\text{c} - \frac{1}{a} \int^\tau a A_\text{c} d\tau' = \sigma_\text{c} + \frac{1}{a} \int^\tau \frac{a}{\mathcal{H}} \, \mathcal{R}^\prime d\tau' \; .
\end{equation}

\noindent \textbf{Uniform Hubble gauge: $\delta \mathcal{H} \equiv - \mathcal{H} A - \psi^\prime \! + \frac{1}{3} \nabla^2 \sigma^\text{S} \! = 0$.} By Eq.~(\ref{eq:theta_pert}), the correction to the local conformal Hubble parameter at linear order is
\begin{equation} \label{eq:deltaH}
\delta \mathcal{H} \equiv \frac{a}{3} \theta - \mathcal{H} = - \mathcal{H} A - \psi^\prime + \frac{1}{3} \nabla^2 \sigma^\text{S} \; ,
\end{equation}
which transforms as $\delta \mathcal{H} \rightarrow \delta \mathcal{H} - ( \mathcal{H}^\prime \! - \! \mathcal{H}^2 ) \xi^0 - \frac{1}{3} \nabla^2 \xi^0$ under the coordinate transformation (\ref{eq:coordtransS}). Therefore, we can define the following gauge-invariant quantities to represent the perturbations in this gauge where $\delta \mathcal{H} = 0$,
\begin{align}
A_\text{H} &\equiv A - \frac{1}{a} \bigg( \frac{a}{( \mathcal{H}^\prime \! - \! \mathcal{H}^2 ) + \frac{1}{3} \nabla^2} \, \delta \mathcal{H} \bigg)^\prime \; , \\
\psi_\text{H} &\equiv \psi + \frac{\mathcal{H}}{( \mathcal{H}^\prime \! - \! \mathcal{H}^2 ) + \frac{1}{3} \nabla^2} \, \delta \mathcal{H} \; , \\
\sigma_\text{H} &\equiv \sigma^\text{S} - \frac{1}{( \mathcal{H}^\prime \! - \! \mathcal{H}^2 ) + \frac{1}{3} \nabla^2} \, \delta \mathcal{H} \; , \\
\delta\phi_\text{H} &\equiv \delta\phi - \frac{\phi^\prime}{( \mathcal{H}^\prime \! - \! \mathcal{H}^2 ) + \frac{1}{3} \nabla^2} \, \delta \mathcal{H} \; .
\end{align}
The curvature perturbation $\psi_\text{H}$ and shear perturbation $\sigma_\text{H}$ in this gauge can be expressed in the comoving gauge as
\begin{align}
\psi_\text{H} &= \frac{( \mathcal{H}^\prime \! - \! \mathcal{H}^2 )}{( \mathcal{H}^\prime \! - \! \mathcal{H}^2 ) + \frac{1}{3} \nabla^2} \Big( - \frac{\mathcal{R}^\prime}{3 c_s^2 \mathcal{H}} + \mathcal{R} \Big) \; , \\
\sigma_\text{H} &= \frac{( \mathcal{H}^\prime \! - \! \mathcal{H}^2 )}{( \mathcal{H}^\prime \! - \! \mathcal{H}^2 ) + \frac{1}{3} \nabla^2} \; \sigma_\text{c} \; . \label{eq:sigmac-to-sigmaH}
\end{align}

\subsection{vector perturbations}

There are no vector contributions to the stress-energy tensor from the scalar field, so the equations of motion are simply
\begin{align}
& \frac{1}{2 a^2} \, \nabla^2 ( F_i^{\,\prime} - S_i ) = 0 \; , \label{eq:G0iv} \\[4pt]
& \frac{1}{a^2} \Big[ ( F_{(i,j)}^{\,\prime} - S_{(i,j)} )^\prime + 2 \mathcal{H} ( F_{(i,j)}^{\,\prime} - S_{(i,j)} ) \Big] = 0 \; , \label{eq:Gijv}
\end{align}
which correspond to the vector modes of the $(^0_{\;i})$ and $(^i_{\;j})$ components of the Einstein equation respectively. The combination
\begin{equation}
\sigma^\text{V}_i \equiv F_i^{\,\prime} - S_i
\end{equation}
is by itself gauge-invariant under the vector mode of the coordinate transformation
\begin{equation}
x^i \rightarrow x^i + \eta^i \; ,
\end{equation}
where $\eta^i_{\; ,i} = 0$. Therefore the vector shear perturbation $\sigma^\text{V}_i$ is the one and only physical degree of freedom, which obeys the equations
\begin{align}
& \nabla^2 \sigma^\text{V}_i = 0 \; , \label{eq:sigmaVec0i} \\[4pt]
& \sigma^\text{V}_i {}^\prime + 2 \mathcal{H} \, \sigma^\text{V}_i = 0 \; . \label{eq:sigmaVecij}
\end{align}

\end{document}